\documentclass[fleqn,11pt]{wlscirep}
\usepackage[utf8]{inputenc}
\usepackage[T1]{fontenc}
\usepackage[numbers,square,sort&compress]{natbib}

\usepackage{amsthm}
\usepackage{bibentry}
\usepackage{color}
\usepackage{epsfig}
\usepackage{framed}
\usepackage{graphics}
\usepackage{hyperref}
\usepackage[left, modulo]{lineno}
\usepackage{lmodern}
\usepackage{multirow}
\usepackage{relsize}
\usepackage{rotating}
\usepackage{subfigure}
\usepackage{subfiles}
\usepackage{url}
\usepackage{verbatim}
\usepackage{wrapfig}
\usepackage{floatrow}
\newfloatcommand{capbtabbox}{table}[][\FBwidth]

\usepackage[noend]{algpseudocode}
\usepackage{algorithm}

\usepackage{graphicx,psfrag,epsf}
\usepackage{enumerate}

\makeatletter
\@addtoreset{equation}{section}   

\makeatother



\usepackage{mycommands2}
\theoremstyle{definition}



\title{Simultaneous Selection of Multiple Important Single Nucleotide Polymorphisms in Familial Genome Wide Association Studies Data}
\author[1,*]{Subhabrata Majumdar}
\author[1]{Saonli Basu}
\author[1]{Matt McGue}
\author[1]{Snigdhansu Chatterjee}
\affil[1]{University of Minnesota Twin Cities, Minneapolis, USA}
\affil[*]{Currently at AI Risk and Vulnerability Alliance. zoom.subha@gmail.com}


\date{}

\begin{abstract}
We propose a resampling-based fast variable selection technique for detecting relevant single nucleotide polymorphisms (SNP) in a multi-marker mixed effect model. Due to computational complexity, current practice primarily involves testing the effect of one SNP at a time, commonly termed as `single SNP association analysis'. Joint modeling of genetic variants within a gene or pathway may have better power to detect associated genetic variants, especially the ones with weak effects. In this paper, we propose a computationally efficient model selection approach---based on the e-values framework---for single SNP detection in families while utilizing information on multiple SNPs simultaneously. To overcome computational bottleneck of traditional model selection methods, our method trains one single model, and utilizes a fast and scalable bootstrap procedure. We illustrate through numerical studies that our proposed method is more effective in detecting SNPs associated with a trait than either single-marker analysis using family data or model selection methods that ignore the familial dependency structure. Further, we perform gene-level analysis in Minnesota Center for Twin and Family Research (MCTFR) dataset using our method to detect several SNPs using this that have been implicated to be associated with alcohol consumption.
\end{abstract}

\begin{document}
\flushbottom
\maketitle

\section{Introduction}

Genome Wide Association Studies (GWAS) have identified a large number of genetic variants associated with complex diseases \cite{ChangEtal13,WheelerBarroso11}. The advent of economical high-throughput genotyping technology enables researchers to scan the genome with millions of Single Nucleotide Polymorphism (SNP)-s, and improvements in computational efficiency in analysis techniques has facilitated parsing through this huge amount of data to detect significant associations \cite{VisscherEtal12}. However, detecting small effects of individual SNPs requires large sample size \cite{ManolioEtal09}. For quantitative behavioral traits such as alcohol consumption, drug abuse, anorexia and depression, variation in genetic effects due to environmental heterogeneity brings in additional noise, further amplifying the issue. This is one of the motivations of performing GWAS on families instead of unrelated individuals, through which the environmental variation can be reduced \cite{BenyaminEtal09,MillerEtal12,VisscherEtal17}. However, association analysis of multiple SNPs while using dependent data with a familial structure and large sample sizes can be computationally very challenging. Thus single SNP association analysis is the standard tool for detecting SNPs, and most family studies tend to have smaller sample size. The MCTFR Study \cite{MillerEtal12} with genome-wide data on identical twins, non-identical twins, biological offspring, adoptees serve as the motivation for our methodology development in this paper. 

A downside of family-based single-SNP methods---such as GRAMMAR~\cite{AulchenkoEtal07} and the association test of \citet{ChenAbecasis07}---is that they do not take into account shared environment effects {\it within} families. They assume that phenotypic similarity among individuals in a family is entirely due to their genetic similarity and not due to the effect of shared environment.
As a result, they tend to lose power when analyzing data where shared environmental effects explain a substantial proportion of the total phenotypic variation (see \cite{McGueEtal13,DeNeveEtal13} for examples). The RFGLS method proposed by \citet{LiEtal11} does take into account genetic and environmental sources of familial similarity and provides fast inference through a rapid approximation of SNP-specific coefficients from a mixed effect model. However it is only able to handle single SNPs at a time.

Single-SNP methods are less effective in detecting SNPs with weak signals \cite{ManolioEtal09}. This is limiting in situations where multiple SNPs are jointly associated with the phenotype \cite{YangEtal12, Ke12, SchifanoEtal12}. Several methods of multi-SNP analysis have been proposed as alternatives. The kernel based association tests \cite{SchifanoEtal12, ChenEtal13, SchaidEtal13, ILazaEtal13} are prominent among such techniques. However, all such methods 
test for whether a group of SNPs is associated with the phenotype of interest {\it as a whole}, and do not prioritize within that group to detect the individual SNPs primarily associated with the trait.
One way to solve this problem is to perform model selection. The methods of \citet{FrommeletEtal12} and \citet{ZhangEtal14} take this approach, and perform SNP selection from a multi-SNP model on GWAS data from \textit{unrelated individuals}. However, they rely on fitting models corresponding to multiple predictor sets, hence are computationally very intensive to implement in a linear mixed effect framework for modeling familial data.


In this paper we propose a fast and scalable model selection technique that fits a single model to a familial dataset, and aims to identify genetic variants with weak signals that are associated with the outcome through joint modelling of multiple variants. We consider only main effects of the variants, but this can be extended to include higher-order interactions. We achieve this by extending the framework of e-values \cite{evalues}, which we discuss in Section~\ref{sec:evalues}. There we present the definition of an e-value, discuss some of its properties, and describe how we generalize the e-values for our scenario.  Broadly, for any estimation method that provides consistent estimates (at a certain rate relative to the sample size) of the vector of parameters, e-values quantify the proximity of the sampling distribution for a restricted parameter estimate to that of the full model estimate in a regression-like setup. A variable selection algorithm using the e-values has the three simple and generic steps. First, fit the full model, i.e. where all predictor effects are being estimated from the data, and use resampling to estimate its e-value. Second, set an element of the full model coefficient estimate to 0 and get an e-value for that predictor using resampling distribution of previously estimated parameters- repeat this for all predictors. Then finally, select predictors that have e-values below a pre-determined threshold.

The above algorithm offers multiple important benefits in the SNP selection scenario. Unlike other model selection methods, only the full model needs to be computed here. It thus offers the user more flexibility in utilizing a suitable method of estimation for the full model. Our method allows for fitting multi-SNP models, thereby accommodating cases of modelling multiple correlated SNPs or closely located multiple causal SNPs simultaneously. Finally, we use the Generalized Bootstrap (GBS)\cite{ChatterjeeBose05} as our chosen resampling technique. Instead of fitting a separate model for each bootstrap sample, GBS computes bootstrap estimates using Monte-Carlo samples from the resampling distribution as weights, and reusing model objects obtained from the full model. Consequently, the resampling step becomes very fast and parallelizable.

In past literature, \citet{VanderweeleDing17} and \citet{VovkWang21} used the term `e-value' in the contexts of sensitivity analysis and multiple testing, respectively. In comparison, the e-values we use~\cite{evalues} {\it evaluate} the relevance of a variable with reference to a statistical model. Going beyond the existing proposal of e-values tied to specific objectives and models, as well as the well-known $p$-values used for hypothesis testing, this e-value is assumption-lean, covers more generic statistical problems---such as including dependent data models---and is expandable to numerous applications, including group feature selection, hypothesis testing, and multiple testing.

\section{Materials and Methods}
\label{sec:modelSection}

\subsection{The MCTFR data}
The familial GWAS dataset collected and studied by Minnesota Center for Twin and Family Research (MCTFR)\cite{LiEtal11, MillerEtal12, McGueEtal13} consists of samples from three longitudinal studies conducted by the MCTFR: (1) the Minnesota Twin Family Study (MTFS)\cite{IaconoEtal99} that covers twins and their parents, (2) the Sibling Interaction and Behavior Study (SIBS)\cite{McGueEtal07} that includes adopted and biological sibling pairs and their rearing parents, and (3) the enrichment study \cite{KeyesEtal09} that extended the MTFS by oversampling 11 year old twins who are highly likely to develop substance abuse. While 9827 individuals completed the phenotypic assessments for participation in the study, after several steps of screening \cite{MillerEtal12} the genotype data
from 7605 Caucasian individuals clustered in 2151 nuclear families were included in our analysis. This consisted of 1109 families where the children are identical twins, 577 families with non-identical twins, 210 families with two adopted children, 162 families with two non-twin siblings, and 93 families where one child is adopted while the other is the biological child of the parents.

DNA samples collected from the subjects were analyzed using Illumina's Human660W-Quad Array, and after standard quality control steps \cite{MillerEtal12}, 527,829 SNPs were retained. Covariates for each sample included age, sex, birth year, generation (parent or offspring), as well as the two-way interactions generation x age, generation x sex, and generation x birth year. Five quantitative phenotypes measuring substance use disorders were studied in this GWAS: (1) Nicotine dependence, (2) Alcohol consumption, (3) Alcohol dependence, (4) Illegal drug usage, and (5) Behavioral disinhibition. The response variables corresponding to these phenotypes are derived from questionnaires using a hierarchical approach based on factor analysis \cite{HicksEtal11}.

A detailed description of the data is available  in \citet{MillerEtal12}. Several studies reported SNPs associated with phenotypes collected in MCTFR study \cite{LiEtal11, McGueEtal13, CoombesBasuMcGue17}. \citet{LiEtal11} used RFGLS to detect association between height and genetic variants through single-SNP analysis, while \citet{McGueEtal13} used the same method to study SNPs influencing the development of all five indicators of behavioral disinhibition mentioned above. \citet{IronsThesis12} focused on the effect of several factors affecting alcohol use in the study population, namely the effects of polymorphisms in the ALDH2 gene and the GABA system genes, as well as the effect of early exposure to alcohols as adolescents to adult outcomes. Finally \citet{CoombesBasuMcGue17} used a bootstrap-based combination test and a sequential score test to evaluate gene-environment interactions for alcohol consumption.

\subsection{Consents and Approvals}
Data were collected through the Minnesota Center for Twin and Family Research (MCTFR). All University of Minnesota and National Institute of Health (NIH) guidelines for human subjects research were followed in the collection and processing of the data. The protocol was approved by the Institutional Review Board (IRB) at the University of Minnesota (protocol \# 0303M45703). Participants aged 18 years and older completed informed consent, while consent was obtained from at least one parent for those participants younger than 18 and the minor participant also assented to participate.



%

\subsection{Statistical model}
\label{subsec:ACE}

We use a Linear Mixed Model (LMM) with three variance components accounting for several potential sources of variation to model effect of SNPs behind a quantitative phenotype. This is known as \textit{ACE model} in the literature \cite{KohlerEtal11}. While the-state-of-the-art focuses on detection of a {\it single variant at a time}, we will incorporate \textit{all} SNPs genotyped within a gene (or group of genes in some cases) as set of fixed effects in a \textit{single model}.

Following standard protocol for family-based GWAS~\cite{ChenAbecasis07,LiEtal11,McGueEtal13}, we assume a data setting of nuclear pedigrees, i.e. that the data consists of observations from individuals of multiple genetically unrelated families, with individuals within a family potentially sharing genetic material. Suppose there are $m$ such families in total, with the $i^{\Th}$ pedigree containing $n_i$ individuals, and the total number of individuals is $n = \sum_{i=1}^m n_i$. Denote by $\bfy_i = (y_{i 1}, \ldots, y_{i n_i})^T $ the quantitative trait values for individuals in that pedigree, while the matrix $\bfG_i \in \BR^{ n_i \times p_g}$ contains their genotypes for a number of SNPs. Let $\bfC_i \in \BR^{ n_i \times p}$ denote the data on $p$ covariates for individuals in the pedigree $i$. Given these, we consider the following model.
\begin{align}
\label{eqn:LMMeqn}
\bfY_i = \alpha + \bfG_i \bfbeta_g + \bfC_i \bfbeta_c + \bfepsilon_i,
\end{align}
with $\alpha$ the intercept term, $\bfbeta_g$ and $\bfbeta_c$ fixed coefficient terms corresponding to the multiple SNPs and covariates, respectively, and $\bfepsilon_i \sim \cN_{n_i} ({\bf 0}, \bfV_i)$ the random error term. To account for the within-family dependency structure, we break up the random error variance into three independent components:
\begin{align}
\label{eqn:partsOfV}
\bfV_i = \sigma_a^2 \bfPhi_i + \sigma_c^2 {\bf 1} {\bf 1}^T + \sigma_e^2 \bfI_{n_i}.
\end{align}

The three components of $\bfV_i$ in \eqref{eqn:partsOfV} model different sources of random variations that can affect the quantitative trait values for individuals in the $i^{\Th}$ pedigree. The first component above is a within-family random effect term to account for shared polygenic effects. The proportion of of genetic material shared between pairs of individuals in a family is represented by elements of the matrix $\bfPhi_i$. Its $(s,t)^{\Th}$ element represents two times the kinship coefficient, which is the probability that two alleles, one randomly chosen from individual $s$ in pedigree $i$ and the other from individual $t$, are `identical by descent', i.e. come from same common ancestor \cite{KohlerEtal11}. Following basic probability, the kinship coefficient of a parent-child pair is 1/4, a full sibling pair or non-identical (or dizygous = DZ) twins is 1/4, and for identical (or monozygous = MZ) twins is 1/2 in a nuclear pedigree. Following this, we can construct the $\bfPhi_i$ matrices for different types of families:
\begin{align*}
\bfPhi_{MZ} = \begin{bmatrix}
1 & 0 & 1/2 & 1/2 \\
0 & 1 & 1/2 & 1/2 \\
1/2 & 1/2 & 1 & 1\\
1/2 & 1/2 & 1 & 1
\end{bmatrix},
\bfPhi_{DZ} = \begin{bmatrix}
1 & 0 & 1/2 & 1/2 \\
0 & 1 & 1/2 & 1/2 \\
1/2 & 1/2 & 1 & 1/2\\
1/2 & 1/2 & 1/2 & 1
\end{bmatrix},
\bfPhi_{Adopted} = \bfI_4.
\end{align*}
for families with parents (indices 1 and 2) and MZ twins, DZ twins, or two adopted children (indices 3 and 4), respectively.

The second variance component $\sigma_c^2 {\bf 1} {\bf 1}^T$ in \eqref{eqn:partsOfV} accounts for shared environmental effect within each pedigree. Traits of each individual in the pedigree are affected by the same amount---a single random draw from $N(0,\sigma_c^2)$---of random variation. The third term in \eqref{eqn:partsOfV} quantifies other sources of variation unique to each individual.

\subsection{Feature Selection with e-values}
\label{sec:evalues}
We extend the recently-proposed framework of e-values \cite{evalues} to select important SNPs in the above gene-level, multi-SNP statistical model. In a general modelling situation where one needs to estimate a set of parameters $\bftheta \in \BR^d$ from data with sample size $n$, a statistical model corresponds to a subset of the full parameter space. In other words, the estimable index set of $\theta$, say $\cS \subseteq \{ 1, \ldots, d \}$ specifies a model. The other indices are set at constant values---typically in model selection literature the constants are set at 0. Note that we are attempting to select important SNPs as described in Section~\ref{subsec:ACE}, thus in our setting $\theta \equiv \beta_g, d \equiv p_g$.

Following the recipe in \citet{evalues}, we obtain coefficient estimates corresponding to model $\cS$ by simply replacing elements of the `full model' estimate $\hat \theta$---i.e. the the coefficient estimate with all possible parameters included---at indices not in $\cS$:
$$
\hat \theta_{\cS} = \left\{ \begin{array}{ll}
\hat \theta_{j} & \text{ for } j \in \cS, \\
0 & \text{ for } j \notin \cS.
\end{array}
\right.
$$

{\it Sampling distribution} is defined as the distribution of a parameter estimate, based on the random data samples used to calculate this estimate. We compare sampling distributions of the above model with the full model, i.e. $[\hat \bftheta_\cS]$ with  $[\hat\bftheta]$ (denoting the distribution of a random variable by $[\cdot]$). For this comparison, we define an \textit{evaluation map} function $E: \BR^d \times \tilde \BR^d \raro [0, \infty)$ that measures the relative position of $\hat \bftheta_\cS$ with respect to $[\hat \bftheta]$. Here $\tilde \BR^d$ is the set of probability measures on $\BR^d$. For any $\bfx \in \BR^d$ and $[ \bfX ] \in \tilde \BR^d$ with a positive definite covariance matrix $\BV \bfX$, we consider the following evaluations functions in this paper:
\begin{align}\label{eqn:EvaluationExamples}
E_1 (\bfx, [ \bfX]) &= \left[ 1 + \left\| (\diag  (\BV \bfX))^{-1/2} \odot (\bfx - \BE \bfX)  \right\|^2 \right]^{-1},\\
E_2 (\bfx, [ \bfX]) &= \exp \left[ - \left\| (\diag  (\BV \bfX))^{-1/2} \odot (\bfx - \BE \bfX)  \right\| \right].
\end{align}
Here $\diag  (\BV \bfX)$ denotes the vector composed of the diagonal entries in $\BV \bfX$, and $\odot$ represents elementwise product, so that $(\diag  (\BV \bfX))^{1/2}$ is the vector of coordinate-wise standard deviation, and $(\diag  (\BV \bfX))^{-1/2} \odot (\bfx - \BE \bfX)$ is a normalized version of $\bfx$. Data depths \cite{tukey75, zuo00} also constitute a broad class of functions that can be used as evaluation maps---as done by \citet{evalues}. In general, any continuous function that is location and scale invariant, and has a few basic convergence properties is a good choice for the evaluation map function (see conditions E1-E4 in Appendix~\ref{subsec:DefineModel}).

\subsubsection{Formulation}
Note that the evaluation map function is defined conditional on a fixed value of $\hat \theta_\cS$. Since $\hat \theta_\cS$ itself has a distribution, so does the evaluation map. We define the \textbf{e-value} as any functional of the evaluation map distribution $\BE_{\cS} \equiv [ E( \hat \theta_\cS, [\hat \theta]) ]$ that can act as a measure of comparison between the sampling distributions of $\hat \theta_\cS$ and $\hat \theta$. For example, Majumdar and Chatterjee~\cite{evalues} took the mean functional of $\BE_{\cS }$ (say $\mu (\BE_{\cS })$) as e-value, and showed that it can be used as a model selection criterion. To this end, non-zero indices (say $\cS_0$) of the true parameter vector $\bftheta_0$ can be recovered through a fast algorithm that has these generic steps:

\begin{table}[h]
\begin{tabular}{p{\linewidth}}
\hline
{\bf Algorithm 1: Model Selection with e-values} \\\hline
1.\quad Obtain the e-value of the full model, say $\mu(\BE_*)$. \\
2.\quad For the $j^{\Th}$ predictor, $j=1,\ldots, p$, consider the model with the $j^{\Th}$ coefficient of $\hat \bftheta$ replaced by 0, and obtain its e-value---denoted by $\mu(\BE_{-j})$. \\
3.\quad Collect the predictors for which $\mu(\BE_{-j}) < \mu(\BE_*)$: denote it by $\hat \cS_0$. This is the estimated set of non-zero coefficients in $\hat \bftheta$.\\\hline
\end{tabular}
\end{table}

\noindent As $n \raro \infty$, the above algorithm provides consistent model selection, i.e. $\BP ( \hat \cS_0 = \cS_0) \raro 1$. In practice we only have one dataset, so it is not possible to access the true sampling distribution of $\hat \theta$ and $\hat \theta_\cS$ to do the above. To this end, we use a fast bootstrap algorithm, called Generalized Bootstrap (GBS)\cite{ChatterjeeBose05}, to obtain approximations of the sampling distributions $[\hat \theta_\cS], [\hat \theta]$, the evaluation map distributions, and the e-values. GBS is dependent of a tuning parameter $\tau_n$ that represents the standard deviation of the synthetic noise introduced by the bootstrap procedure. Intermediate values of $\tau_n$, such that $\tau_n/n \rightarrow \infty$, result in model selection consistency as described above.

\subsubsection{Quantile e-values}
When true signals are weak, the above method of variable selection leads to very conservative estimates of non-zero coefficient indices, i.e. a large number of false positives in a sample setting. This happens because the true values leave-one-covariate-out e-values for variables that correspond to small but non-zero coefficients in $\theta_0$ (hence weak signal) fall too close to the full model e-value. Consequently, when these e-values are estimated from randomly sampled data, simply by random chance their values can be slightly less than the full model e-value estimate.

\begin{figure}
\centering
\includegraphics[width=.49\textwidth]{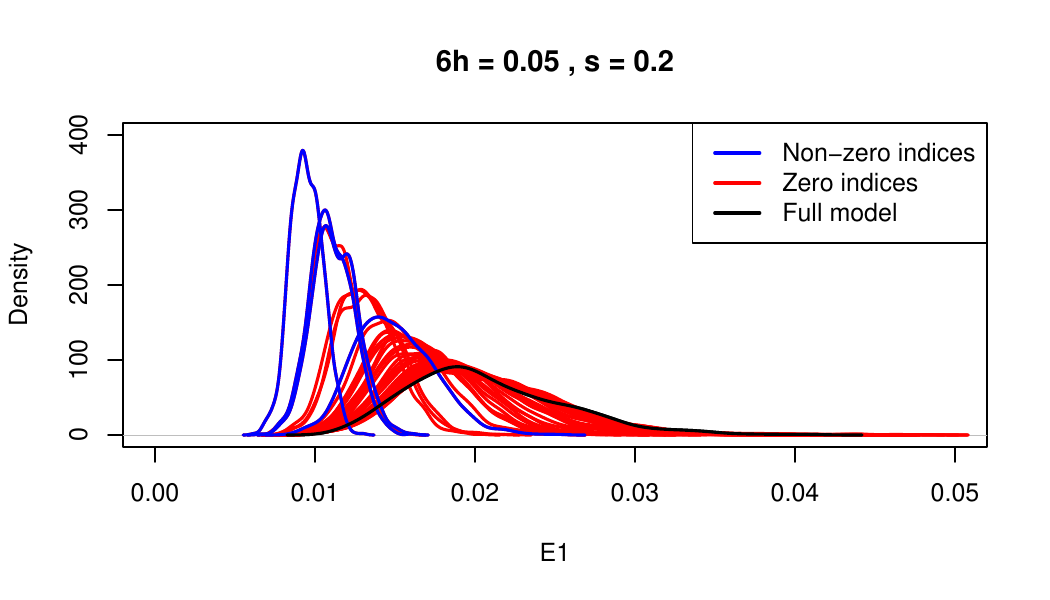}
\includegraphics[width=.49\textwidth]{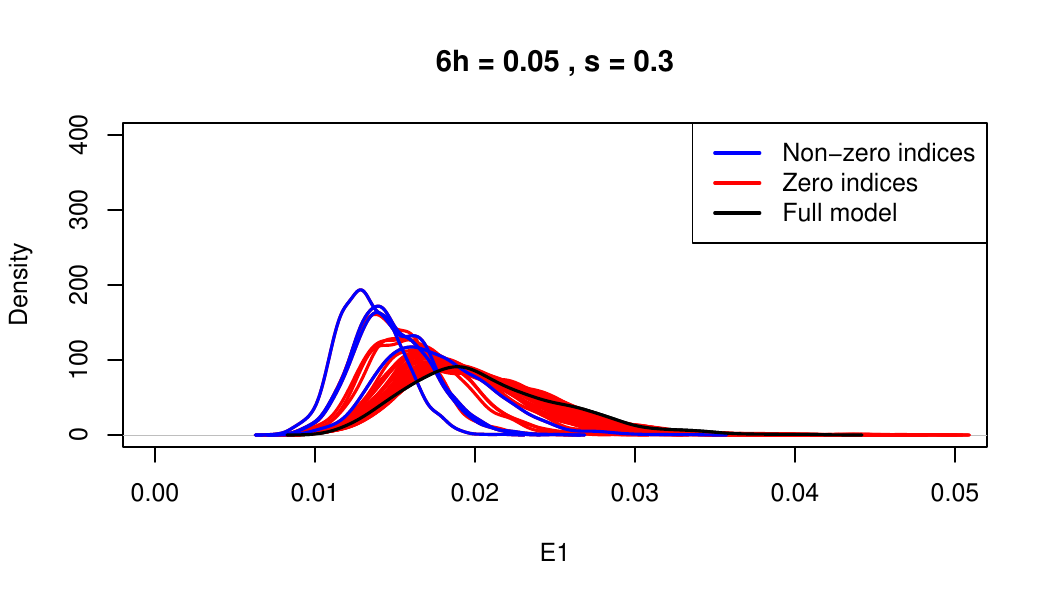}\\
\includegraphics[width=.49\textwidth]{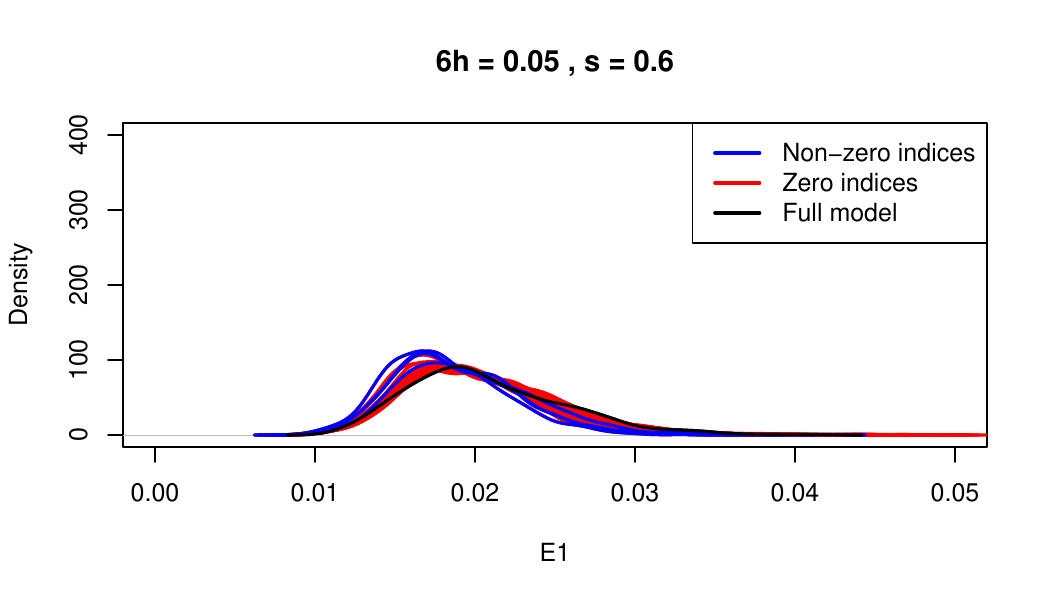}
\includegraphics[width=.49\textwidth]{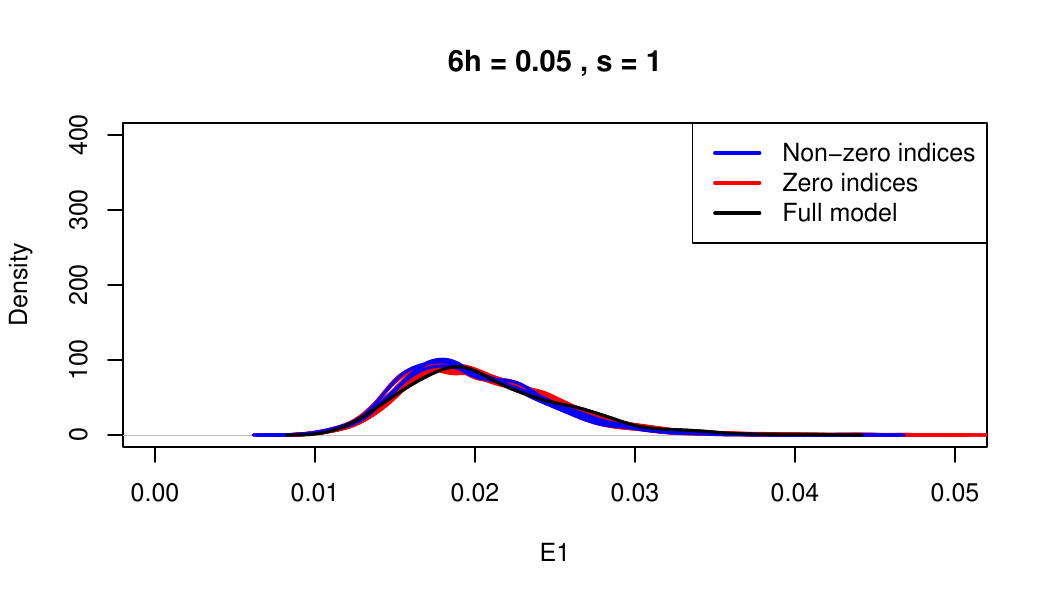}
\caption{Density plots of bootstrap approximations for $\BE_{* }$ and $\BE_{-j}$ for all $j$ in simulation setup, with $s = 0.2, 0.3, 0.6, 1$.}
\label{fig:figSmallhSim}
\end{figure}

Figure \ref{fig:figSmallhSim} demonstrates this phenomenon in our setup, where we would like to estimate non-zero elements of the fixed effect coefficient vector $\beta_g$ in the model \eqref{eqn:LMMeqn}, i.e. $\beta_g \equiv \theta$. Here we analyze data on 250 families with monozygotic twins, each individual being genotyped for 50 SNPs. Four of these 50 SNPs are causal: each having a heritability of $h/6\%$ with respect to the total error variation present. The four panels show density plots of $\BE_{-j}$ for $j = 1, \ldots, p$, as well as $\BE_{* }$: estimated based on resampling schemes with four different values of the standard deviation parameter $s \equiv s_n = \tau_n / \sqrt n $. Focusing on where the central regions of the evaluation map distributions are, we notice that for smaller values of $s$ there is quite a bit of overlap along the bootstrap estimates of $\BE_{-j}$ for causal vs. non-causal SNPs. On the other hand, for large values of $s$ all the density plots become essentially the same as the full model.

However, notice that the evaluation map distributions for non-zero vs. zero indices have different tail behaviors at smaller values of $s$. In the Appendix we show that the means and tail quantiles for $\BE_{-j}$ and $\BE_{*}$ asymptotically converge to different limits (Theorems \ref{thm:meanThm} and \ref{thm:quantileThm}), and these limits are well-approximated with a GBS scheme having small standard deviation $s$ (Theorem~\ref{thm:bootThm}). A potential reason for the different tail behaviors we see above is that the convergence at tail quantiles happens at a faster rate than convergence at the means.

Consequently, instead of comparing means of the distributions, comparing a suitable tail quantile across the distributions is more likely to provide a better separation of non-zero vs. zero indices. For this reason we use tail quantiles as e-values.



When the $q^\text{th}$ quantile (denoted by $c_q$, $q \in (0,1)$) is taken as the e-value instead of the mean, we set a lower detection threshold than the same functional on the full model, i.e. choose all $j$ such that
\begin{align}\label{eqn:cq}
    c_q (\BE_{-j}) < t c_q (\BE_{*}), \quad 0 < t < 1,
\end{align}
to be included in the model. The optimal choice of $q$ and $t$ depends on factors such as 
specifications of the statistical model, sample size, and degree of sparsity of parameters in the data generating process. We demonstrate this point through our experiments in Section~\ref{sec:ExpSection}. For $q$, we take the conservative route by only flagging a SNP as `detected' if Eq.~\eqref{eqn:cq} holds for {\it all} $q \in \{0.5, 0.6, 0.7, 0.8, 0.9\}$. This approach leads to a tradeoff between the true positive and true negative SNP detections rates for different values of $t$. We demonstrate this fact through synthetic data experiments (Section~\ref{sec:SimSection}), and choose the best $t$ that minimizes prediction error on a holdout sample in the MCTFR data analysis (Section~\ref{sec:DataSection}).

\section{Experiments}
\label{sec:ExpSection}

We now evaluate the performance of the above formulation of quantile e-values in through on synthetic data, as well as the MCTFR Twin Studies dataset.

\subsection{Synthetic Data}
\label{sec:SimSection}
Consider the model in (\ref{eqn:LMMeqn}) with no environmental covariates and familes with MZ twins. We take a total of $p_g = 50$ SNPs, and generate the SNP matrices $\bfG_i$ in correlated blocks of 6, 4 ,6, 4 and 30 to simulate correlation among SNPs in the genome. We set the correlation between two SNPs inside a block at 0.7, and consider the blocks to be uncorrelated. For each parent we generate two independent vectors of length 50 with the above correlation structure, and entries within each block being 0 or 1 following Bernoulli distributions with probabilities 0.2, 0.4, 0.4, 0.25 and 0.25 (Minor Allele Frequency or MAF) for SNPs in the 5 blocks, respectively. The genotype of a person is then determined by taking the sum of these two vectors: thus entries in $\bfG_i$ can take the values 0, 1 or 2. Finally we set the common genotype of the twins by randomly choosing one allele vector from each of the parents and taking their sum.


We repeat the above process for $m=250$ families. In GWAS generally each associated SNP explains only a small proportion of the overall variability of the trait. To reflect this in our simulation setup, we assume that the first entries in each of the first four blocks above are causal, and each of them explains $h/(\sigma_a^2+\sigma_c^2+\sigma_e^2) \%$ of the overall variability. The term $h$ is known as the \textit{heritability} of the corresponding SNP. The value of the non-zero coefficient in $k$-th block: $k = 1, ..., 4$, say $\beta_k$ is calculated using the formula:
\begin{align}
\beta_k = \sqrt{ \frac{h}{100 (\sigma_a^2+\sigma_c^2+\sigma_e^2). 2 \text{MAF}_k (1 - \text{MAF}_k) }}.
\end{align}
We fix the following values for the error variance components: $\sigma_a^2 = 4, \sigma_c^2 = 1, \sigma_e^2 = 1$, and generate pedigree-wise response vectors $\bfy_1, \ldots, \bfy_{250}$ using the above setup. To consider different SNP effect sizes, we repeat the above setup for $h \in \{10, 7, 5, 3, 2, 1, 0 \}$, generating 1000 datasets for each value of $h$.

\paragraph{Competing methods}
We compare our e-value based approach using the evaluation maps $E_1$ and $E_2$ in (\ref{eqn:EvaluationExamples}) with two groups of methods:

{\it (1) Model selection on linear model:} Here we ignore the dependency structure within families by training linear models on the simulated data and selecting SNPs with non-zero effects by backward deletion using a modification of the BIC called mBIC2. This has been showed to give better results than single-SNP analysis in a GWAS with unrelated individuals \cite{FrommeletEtal12} and provides approximate False Discovery Rate (FDR) control~\cite{BogdanEtal11}.

{\it (2) Single-marker mixed model:} We train single-SNP versions of (\ref{eqn:LMMeqn}) using a fast approximation of the Generalized Least Squares procedure (named Rapid Feasible Generalized Least Squares or RFGLS~\cite{LiEtal11}), obtain marginal $p$-values from corresponding $t$-tests and use two methods to select significant SNPs: the Benjamini-Hochberg (BH) procedure, as well as the Local FDR method~\cite{lfdr} (LFDR).

For mBIC2 and BH, we choose a conservative FDR level of 0.05 guided by choices in existing work~\cite{KorthauerEtal2019,StoreyTibshirani2003}. Higher FDR values lead to marginal increase in true positive rate but sharp decreases in true negative rate (see definitions of these metrics below). LFDR tends to be much more conservative than global FDR procedures, so we repeated its simulations for a range of FDR values in $\{0.01, 0.05, 0.1, 0.2, 0.3, 0.4, 0.5\}$. Table~\ref{table:SNPSimTable0} shows the performance metrics at FDR = $0.4$, the highest value where the true negative rate is at least as much as the lowest true negative across all settings of e-values as described below.

With the e-value being the $q^\text{th}$ quantile of the evaluation map distribution, we set the detection threshold value at the $t^\text{th}$ multiple of $q$ for some $0 < t < 1$. This means all indices $j$ such that the $q^\text{th}$ quantile of the bootstrap approximation of $\BE_{-j} $ is less than the $tq^\text{th}$ quantile of the bootstrap approximation of $\BE_{*}$ get selected as the set of active predictors. To enforce stricter control on the selected set of SNPs we repeat this for $q \in \{ 0.5, 0.6, 0.7, 0.8, 0.9 \}$, and take the SNPs that get selected for \textit{all} values of $q$ as the final set of selected SNPs. Guided by empirical experiments, we chose values of $t$ for $E_1$ and $E_2$ that clearly demonstrate the tradeoff of TP/TN (or RTP/RTN) rates for the e-values.

Since the above procedure depends on the bootstrap standard deviation parameter $s$, we repeat the process for $s  \in \{ 0.3, 0.15, \ldots, 0.95, 2 \}$, and take as the final estimated set of SNPs the SNP set $\hat \cS_t (s)$ that minimizes fixed effect prediction error (PE) on an independently generated test dataset $\{ (\bfy_{test,i}, \bfG_{test,i}), i = 1, \ldots, 250 \}$ from the same setup above:
\begin{align*}
& \text{PE}_t (s)  = \sum_{i=1}^{250} \sum_{j=1}^4 \left( y_{test,ij} - \bfg_{test,ij}^T \hat \bfbeta_{\hat \cS_t (s)} \right)^2; \\
& \hat \cS_t = \argmin_s \text{PE}_t (s)
\end{align*}

\paragraph{Metrics}
We use the following metrics to evaluate each method we implement: (1) True Positive (TP), which is the proportion of causal SNPs detected;  (2) True Negative (TN), which is the proportion of non-causal SNPs undetected;  (3) Relaxed True Positive (RTP), which is the: proportion of detecting any SNP in each of the 4 blocks with causal SNPs, i.e. for the selected index set by some method $m$, say $\hat \cS_m$,
$$
\text{RTP} ( \hat \cS_m ) = \frac{1}{4} \sum_{i=1}^4 \BI ( \text{Block } i \cap \hat \cS_m \neq \emptyset ),
$$
and finally (4) Relaxed True Negative (RTN), which is the proportion of SNPs in block 5 undetected. We consider the third and fourth metrics to cover situations in which the causal SNP is not detected itself, but highly correlated SNPs with the causal SNP are. This is common in GWAS \cite{FrommeletEtal12}. We average the above proportions over 1000 replications, and repeat the process for two different ranges of $t$ for $E_1$ and $E_2$.


\paragraph{Results}
\begin{table}[t]
\centering
\scalebox{1}{
    \begin{tabular}{c|l|lllllll}
    \hline
    \multicolumn{2}{c|}{Method}          & $h = 10$    & $h = 7$     & $h = 5$     & $h = 3$     & $h = 2$     & $h = 1$     & $h = 0$     \\ \hline
    \multicolumn{2}{c|}{mBIC2}           & 0.79/0.99 & 0.59/0.99 & 0.41/0.99 & 0.2/0.99  & 0.11/0.99 & 0.05/0.99 & -/0.99 \\
    \multicolumn{2}{c|}{RFGLS+BH}        & 0.95/0.92 & 0.82/0.95 & 0.62/0.97 & 0.29/0.98 & 0.14/0.99 & 0.04/1    & -/1       \\     
    \multicolumn{2}{c|}{RFGLS+LFDR}      & 0.54/0.99 & 0.46/0.99 & 0.39/0.99 &  0.29/0.99 &  0.23/1    &  0.15/1    &  -/0.96    \\ \hline
    ~        & $t = \exp(-1)$ & 0.95/0.98 & 0.87/0.97 & 0.74/0.97 & 0.47/0.97 & 0.28/0.97 & 0.12/0.98 & -/0.99 \\
    ~        & $t = \exp(-2)$ & 0.94/0.98 & 0.85/0.98 & 0.69/0.98 & 0.43/0.98 & 0.25/0.98 & 0.09/0.99 & -/0.99 \\
    $E_1$    & $t = \exp(-3)$ & 0.94/0.99 & 0.82/0.98 & 0.65/0.98 & 0.37/0.99 & 0.2/0.99  & 0.07/0.99 & -/1       \\
    ~        & $t = \exp(-4)$ & 0.92/0.99 & 0.79/0.99 & 0.61/0.99 & 0.32/0.99 & 0.17/0.99 & 0.06/1    & -/1       \\
    ~        & $t = \exp(-5)$ & 0.9/0.99  & 0.75/0.99 & 0.55/0.99 & 0.26/1    & 0.13/1    & 0.04/1    & -/1       \\ \hline
    ~        & $t = 0.8$    & 0.97/0.98 & 0.9/0.97  & 0.79/0.96 & 0.54/0.96 & 0.34/0.97 & 0.15/0.98 & -/0.99 \\
    ~        & $t = 0.74$   & 0.96/0.98 & 0.88/0.97 & 0.75/0.97 & 0.48/0.97 & 0.29/0.98 & 0.12/0.98 & -/0.99 \\
    $E_2$    & $t = 0.68$   & 0.95/0.99 & 0.87/0.98 & 0.72/0.98 & 0.45/0.98 & 0.26/0.98 & 0.1/0.99  & -/0.99 \\
    ~        & $t = 0.62$   & 0.95/0.99 & 0.84/0.98 & 0.68/0.98 & 0.4/0.99  & 0.22/0.99 & 0.09/0.99 & -/0.99    \\
    ~        & $t = 0.56$   & 0.94/0.99 & 0.82/0.99 & 0.65/0.99 & 0.36/0.99 & 0.19/0.99 & 0.07/1    & -/1       \\
    ~        & $t = 0.5$    & 0.92/0.99 & 0.79/0.99 & 0.6/0.99  & 0.31/0.99 & 0.16/1    & 0.05/1    & -/1       \\ \hline
    \end{tabular}
}    
\scalebox{1}{
    \begin{tabular}{c|l|lllllll}
    \hline
    \multicolumn{2}{c|}{Method}          & $h = 10$    & $h = 7$     & $h = 5$     & $h = 3$     & $h = 2$     & $h = 1$     & $h = 0$     \\ \hline
    \multicolumn{2}{c|}{mBIC2}           & 0.84/0.99 & 0.66/0.99 & 0.48/0.99 & 0.26/0.99 & 0.16/0.99 & 0.08/0.99 & -/0.98 \\
    \multicolumn{2}{c|}{RFGLS+BH}        & 0.96/0.99 & 0.83/0.99 & 0.64/0.99 & 0.32/0.99 & 0.16/1    & 0.05/1    & -/1    \\
    \multicolumn{2}{c|}{RFGLS+LFDR}      &  0.55/0.99 &  0.47/0.99 &  0.42/0.99 &  0.37/0.99 &  0.35/1    & 0.31/1    & -/0.97    \\ \hline
    ~        & $t = \exp(-1)$ & 0.95/0.98 & 0.87/0.97 & 0.75/0.97 & 0.5/0.97  & 0.32/0.98 & 0.15/0.98 & -/0.98 \\
    ~        & $t = \exp(-2)$ & 0.94/0.99 & 0.85/0.98 & 0.71/0.98 & 0.45/0.98 & 0.28/0.98 & 0.12/0.99 & -/0.98 \\
    $E_1$    & $t = \exp(-3)$ & 0.94/0.99 & 0.83/0.99 & 0.67/0.99 & 0.39/0.99 & 0.22/0.99 & 0.09/0.99 & -/0.99    \\
    ~        & $t = \exp(-4)$ & 0.92/0.99 & 0.8/0.99  & 0.62/0.99 & 0.33/0.99 & 0.18/0.99 & 0.07/1    & -/1    \\
    ~        & $t = \exp(-5)$ & 0.9/0.99  & 0.75/0.99 & 0.56/0.99 & 0.27/1    & 0.14/1    & 0.05/1    & -/1    \\ \hline
    ~        & $t = 0.8$    & 0.97/0.98 & 0.91/0.97 & 0.8/0.96  & 0.57/0.96 & 0.38/0.97 & 0.2/0.98  & -/0.97 \\
    ~        & $t = 0.74$   & 0.96/0.98 & 0.89/0.98 & 0.76/0.97 & 0.51/0.97 & 0.33/0.98 & 0.15/0.98 & -/0.98 \\
    $E_2$    & $t = 0.68$   & 0.95/0.99 & 0.87/0.98 & 0.73/0.98 & 0.48/0.98 & 0.29/0.98 & 0.12/0.99 & -/0.98 \\
    ~        & $t = 0.62$   & 0.95/0.99 & 0.85/0.99 & 0.69/0.98 & 0.42/0.99 & 0.24/0.99 & 0.11/0.99 & -/0.99    \\
    ~        & $t = 0.56$   & 0.94/0.99 & 0.83/0.99 & 0.66/0.99 & 0.38/0.99 & 0.2/0.99  & 0.08/0.99 & -/0.99    \\
    ~        & $t = 0.5$    & 0.92/0.99 & 0.79/0.99 & 0.61/0.99 & 0.32/0.99 & 0.17/1    & 0.06/1    & -/1    \\ \hline
    \end{tabular}}
\caption{(Top) Average True Positive (TP)/ True Negative (TN) rates for mBIC2, RFGLS+BH and the e-values method with $E_1$ and $E_2$ as evaluation maps and different values of $t$ over 1000 replications, and (Bottom) Average Relaxed True Positive (RTP) and Relaxed True Negative (RTN) rates}
\label{table:SNPSimTable0}
\end{table}

We present the simulation results in table \ref{table:SNPSimTable0}. For all heritability values, applying mBIC2 on linear models performs poorly compared to applying RFGLS and then correcting for multiple testing. This is expected because the linear model ignores within-family error components.

Our method works better than the two competing methods for detecting true signals across different values of $h$: the average TP rate going down slowly than other methods across the majority of choices for $t$. All the competing methods (mBIC2, RFGLS+BH, RFGLS+LFDR) have very high true negative detection rates, which is matched by our method for higher values of $q$. Since all reduced model distributions reside on the left of the full model distribution, we expect the variable selection process to turn more conservative at lower values of $t$. This effect is more noticeable for lower $q$, indicating that the right tails of evaluation map distributions are more useful for this purpose. Finally for $h=0$, we report only TN and RTN values since no signals should ideally be detected. Note also the fact that we report the performance metrics for $E_1, E_2$ considering a very conservative selection process: we only mark a SNP $j$ as `detected' if $c_q (\BE_{-j}) < t c_q (\BE_{*})$ for {\it all} $q \in \{ 0.5, 0.6, 0.7, 0.8, 0.9 \}$. We experimentally observed that relaxing this condition leads to higher strict TP rates but lower strict TN.

RTP performances for all methods are better than the corresponding TP/TN performances. However, for mBIC2 this seems to be due to detecting SNPs in the first four blocks by chance since for $h=0$ its RTN is less than TN. Also $E_2$ seems to perform slightly better than $E_1$, in the sense that it yields a higher TP (or RTP) while having the same TN (or RTN) rates.

Among competing methods, it is interesting to notice that LFDR performs better than the other two for small signal values ($h \leq 3$), but worse at higher $h$. This speaks to the strengths of LFDR in low-signal situations. On the other hand, LFDR is calculated using density estimates of the null and non-null statistic distributions. Since there are only 50 SNPs in our simulation setting (and even less in the real data setting), the resulting instability is a potential reason for its low performance at high values of $h$.

\subsection{Analysis of the MCTFR data}
\label{sec:DataSection}

We now apply the above methods on SNPs from the MCTFR dataset. We assume a nuclear pedigree structure, and for simplicity only analyze pedigrees with MZ and DZ twins. After setting aside samples with missing response variables, we end up with 1019 such 4-member families. We look at the effect of genetic factors behind the response variable pertaining to the amount of alcohol consumption, which is highly heritable in this dataset according to previous studies \cite{McGueEtal13}. We analyze SNPs inside some of the most-studied genes with respect to alcohol abuse: GABRA2, ADH1A, ADH1B, ADH1C, ADH4-ADH7, SLC6A3, SLC6A4, OPRM1, CYP2E1, DRD2, ALDH2, and COMT \cite{CoombesThesis16} through separate gene-level models. None of the ADH genes contained a sufficient number of SNPs to justify analysis individually, so we pooled SNPs across all 7 ADH genes for analysis.
We include sex, birth year, age and generation (parent or offspring) of individuals as covariates to control for their potential effect.

For model selection we use $E_2$ as the evaluation function because of its slighty better performance in the simulations. For each gene-level model, We train the LMM in (\ref{eqn:LMMeqn}) on 75\% of randomly selected families, perform our e-values procedure for $s = 0.2, 0.4, \ldots, 2.8, 3, t = 0.1, 0.15, \ldots, 0.75, 0.8$; and select the set of SNPs that minimizes fixed effect prediction error on the data from the other 25\% of families over this grid of $(s,t)$. Note that we consider a wider range of $t$ than in the simulations. This is because of the fact that instead of demonstrating the tradeoff of true positive and true negative rates, our objective here is to actually choose a set of SNPs. For the competing methods, we set FDR levels at 0.05 for mBIC2 and RFGLS+BH, while choose the level that minimizes fixed effect prediction error on the same holdout data as above for RFGLS+LFDR.

\begin{table}[t]
    \begin{tabular}{l|l|llll}
    \hline
    Gene   & Total no. & \multicolumn{3}{l}{No. of SNPs detected by }\\\cline{3-6}
    & of SNPs & e-value & RFGLS+BH & RFGLS+LFDR & mBIC2 \\\hline
    GABRA2 & 11        & 5       & 0     & 1     & 0     \\
    ADH    & 44        & 3       & 1     & 1     & 0     \\
    OPRM1  & 47        & 25      & 1     & 0     & 0     \\
    CYP2E1 & 9         & 5       & 0     & 0     & 0     \\
    ALDH2  & 6         & 5       & 0     & 1     & 1     \\
    COMT   & 15        & 14      & 0     & 1     & 0     \\
    SLC6A3 & 18        & 4       & 0     & 1     & 0     \\
    SLC6A4 & 5         & 0       & 0     & 1     & 0     \\
    DRD2   & 17        & 0       & 0     & 0     & 1     \\\hline
    \end{tabular}
    \caption{Table of analyzed genes and number of detected SNPs in them by the three methods}
    \label{table:genetable}
\end{table}

\begin{table}[t]
    \begin{tabular}{l|p{3in}|p{2in}}
    \hline
    Gene      & Detected SNPs with                              & Reference for \\
    & known associations & associated SNP                              \\\hline
    GABRA2    & rs1808851, rs279856: close to rs279858                         & \citet{CuiEtal12}                                                    \\\hline
    ADH genes & rs17027523: 20kb upstream of rs1229984 & Multiple studies (\url{https://www.snpedia.com/index.php/Rs1229984}) \\\hline
    OPRM1     & rs12662873: 1 kb upstream of rs1799971                            & Multiple studies (\url{https://www.snpedia.com/index.php/Rs1799971}) \\\hline
    CYP2E1    & rs9419624: 600b downstream of rs4646976; rs9419702: 10kb upstream of rs4838767 & \citet{LindEtal12}\\\hline
    ALDH2     & rs16941437: 10kb upstream of rs671                               & Multiple studies (\url{https://www.snpedia.com/index.php/Rs671}) \\\hline
    COMT      & rs4680, rs165774                                               & \citet{VoiseyEtal11}                                                                               \\\hline
    SLC6A3    & rs464049                                                       & \citet{HuangEtal17}\\\hline
    \end{tabular}
    \caption{Table of detected SNPs with known references}
    \label{table:genetable2}
\end{table}

As seen in Table~\ref{table:genetable}, our e-value based technique detects a much higher number of SNPs than the two competing methods. Our method selects all but one SNP in the genes ALDH2 and COMT. These are small genes of size 50kb and 30kb, respectively, thus SNPs within them have more chance of being in high Linkage Disequilibrium (LD). On the other hand, it does not select any SNPs in SLC6A4 and DRD2. Variants of these genes are known to interact with each other and are jointly associated with multiple behavioral disorders \cite{KarpyakEtal10, WangEtal14}.

A number of SNPs we detect (or SNPs situated close to them) have known associations with alcohol-related behavioral disorders. We summarize this in Table~\ref{table:genetable2}. Prominent among them are rs1808851 and rs279856 in the GABRA2 gene, which are at perfect LD with rs279858 in the larger, 7188-individual version of our twin studies dataset \cite{IronsThesis12}. This SNP is the marker in GABRA2 that is most frequently associated in the literature with alcohol abuse \cite{CuiEtal12}, but was not genotyped in our sample. A single SNP RFGLS analysis of the same twin studies data that used Bonferroni correction on marginal $p$-values missed the SNPs we detect \cite{IronsThesis12}: highlighting the advantage of our approach. In the Appendix, we give a gene-wise discussion of associated SNPs (Appendix~\ref{sec:DataSection}), as well as information on all SNPs (Appendix~\ref{sec:SnpSection}).

\setcounter{figure}{1}
\begin{figure}
\begin{center}

\begin{tabular}{c}
		\includegraphics[height=.4\textwidth]{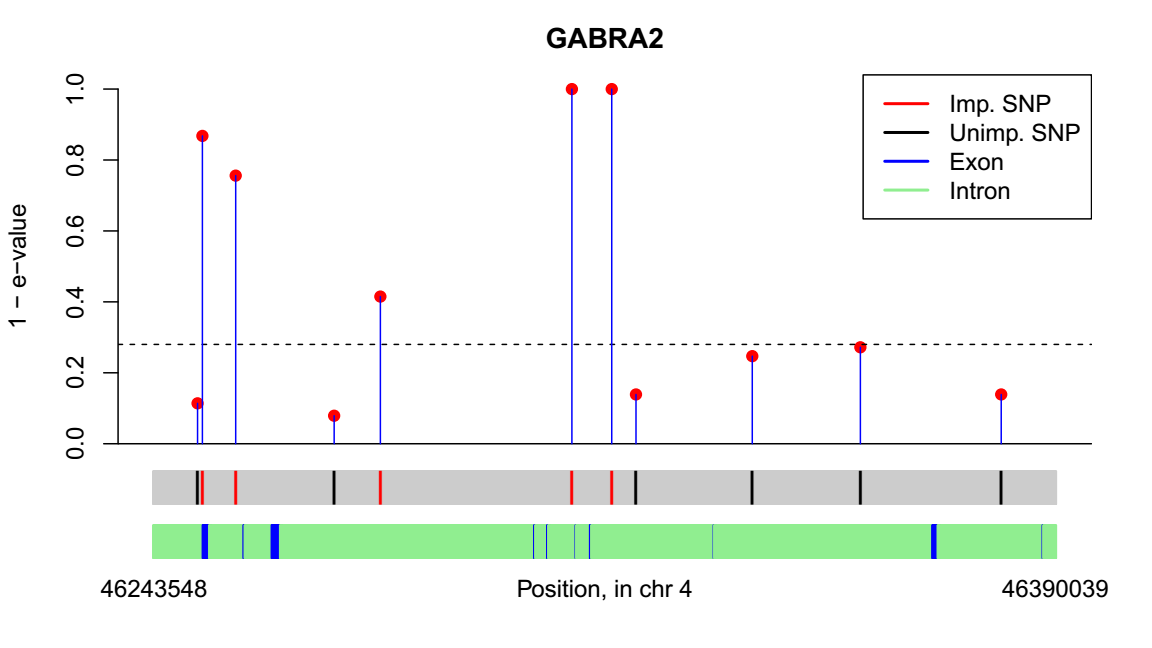}\\
		(a)\\
		\includegraphics[height=.4\textwidth]{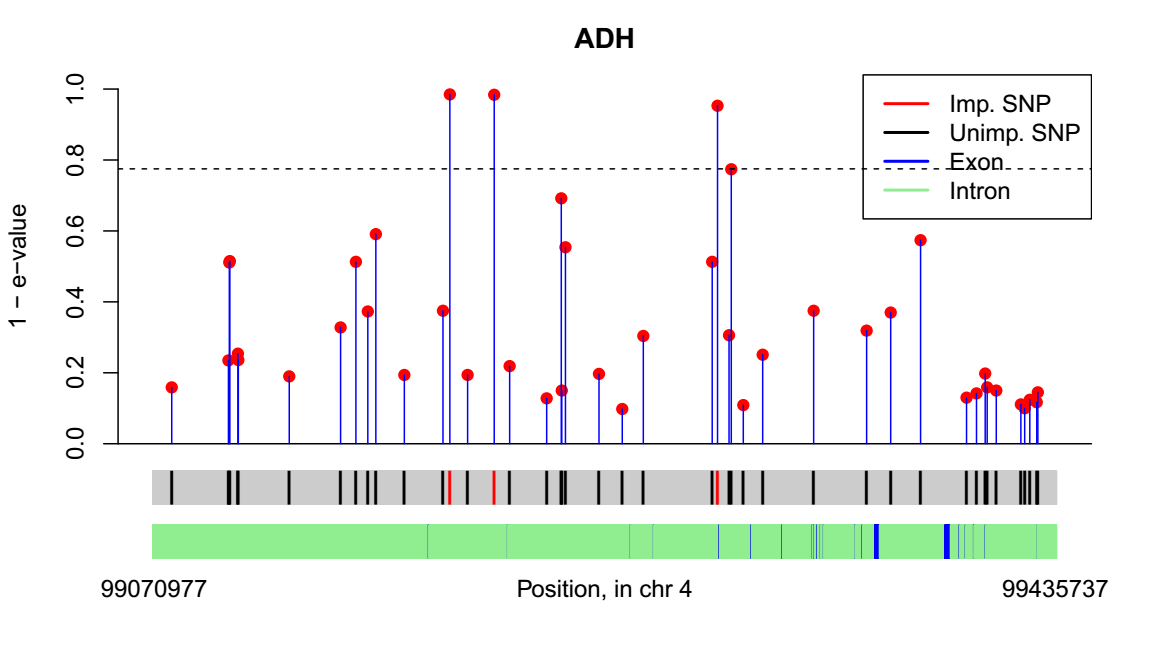} \\
		(b)\\	
		\includegraphics[height=.4\textwidth]{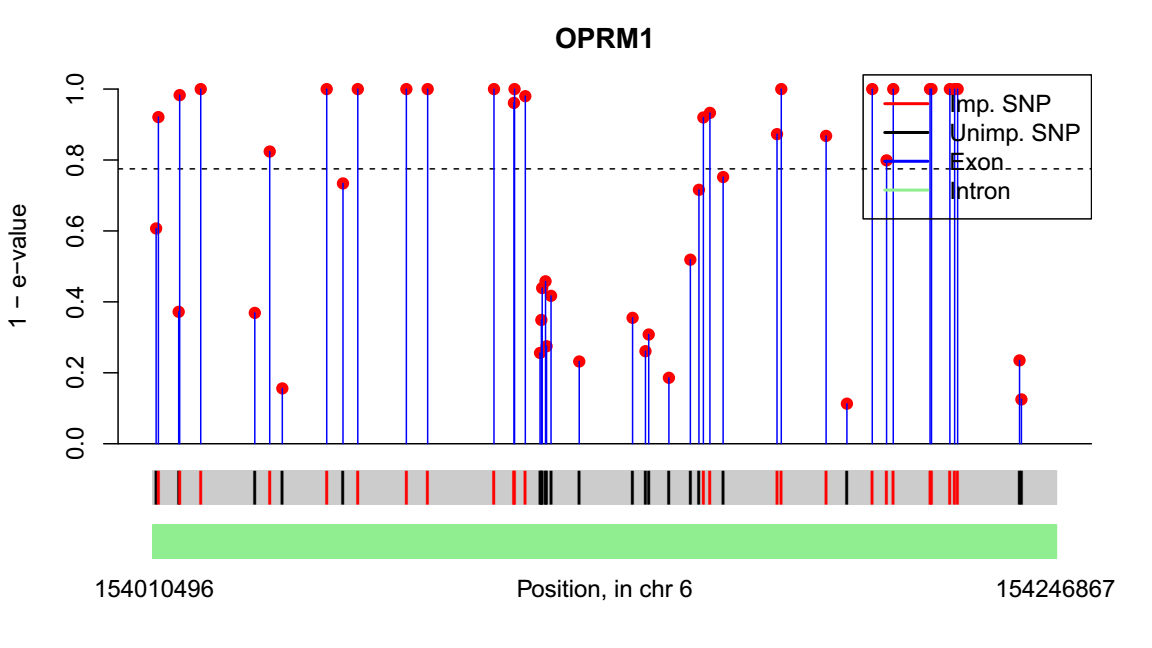}\\
		(c)\\	
\end{tabular}

\caption{Plot of e-values for genes analyzed: (a) GABRA2, (b) ADH1 to ADH7, (c) OPRM1. For ease of visualization, $1 - e$-values are plotted in the y-axis.}
\label{fig:geneplot1}

\end{center}
\end{figure}

\begin{figure}
\begin{center}

\begin{tabular}{c}
		\includegraphics[height=.4\textwidth]{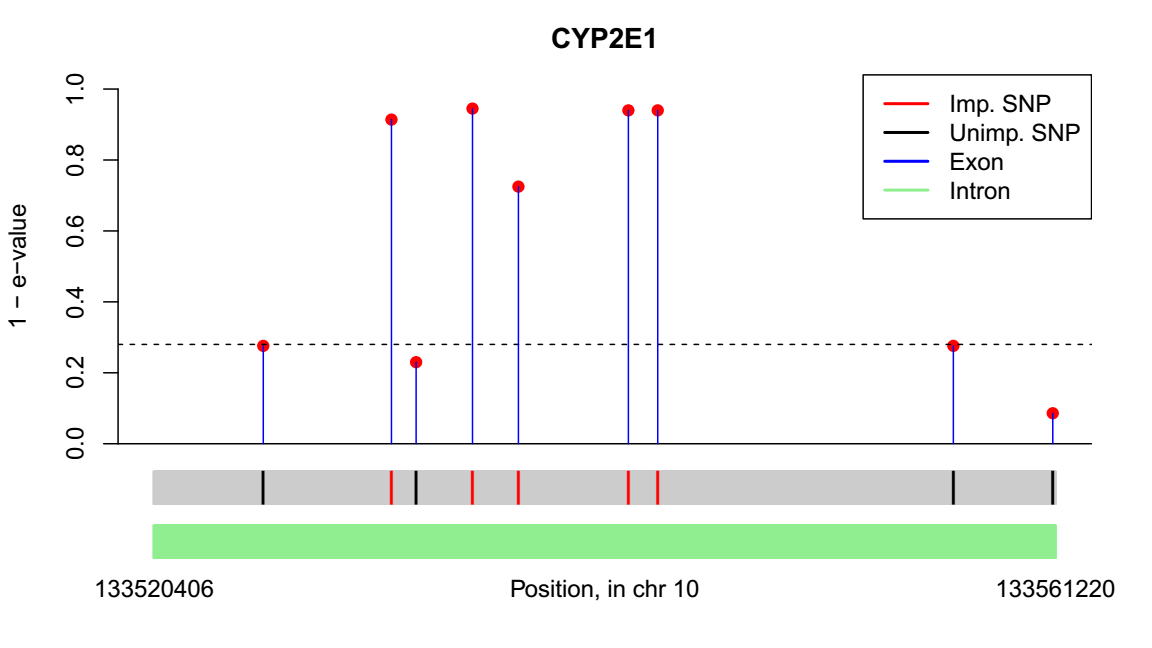}\\
		(d)\\
		\includegraphics[height=.4\textwidth]{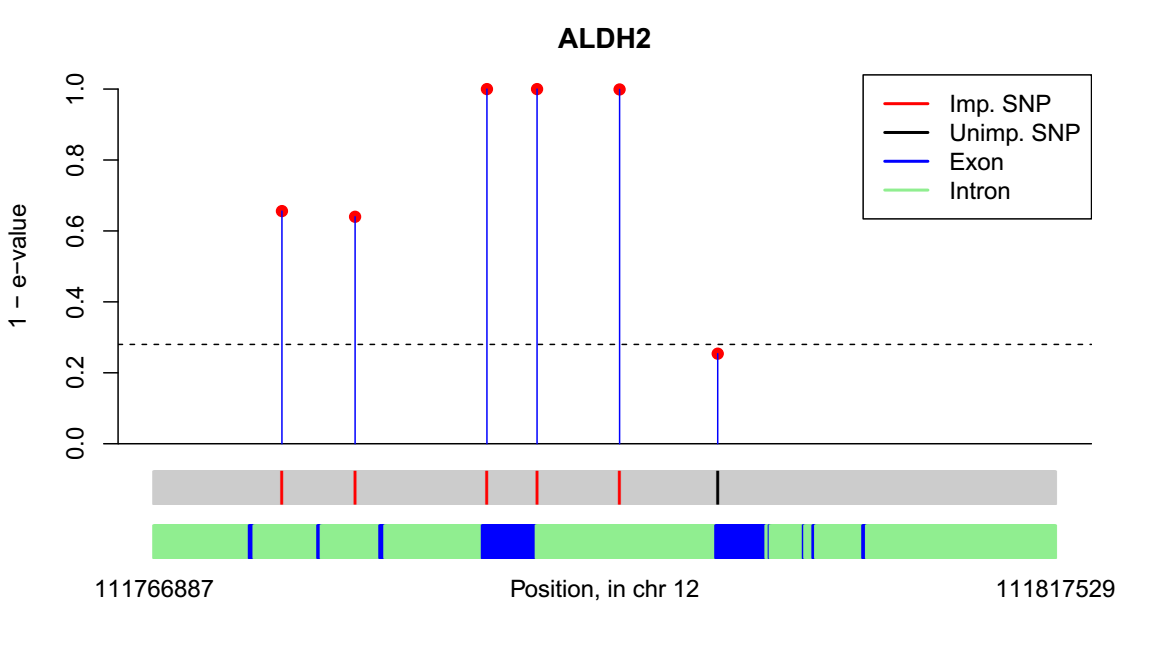} \\
		(e)\\	
		\includegraphics[height=.4\textwidth]{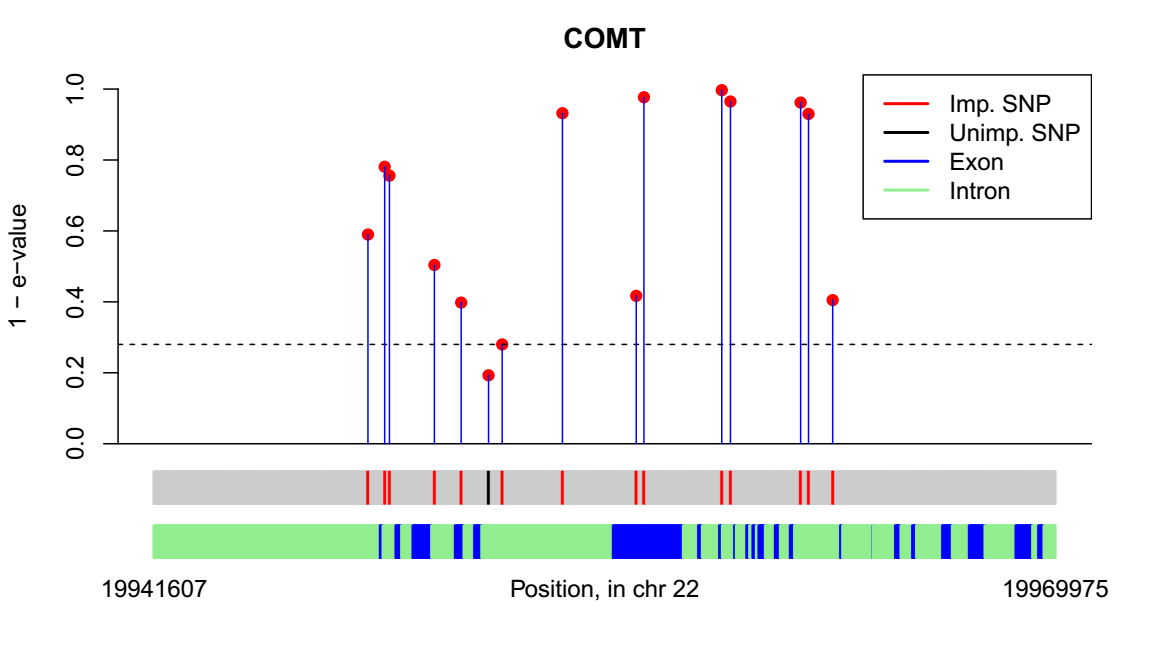}\\
		(f)\\	
\end{tabular}

\caption{Plot of e-values for genes analyzed: (d) CYP2E1, (e) ALDH2, (f) COMT}
\label{fig:geneplot2}

\end{center}
\end{figure}

\begin{figure}
\begin{center}

\begin{tabular}{c}
		\includegraphics[height=.4\textwidth]{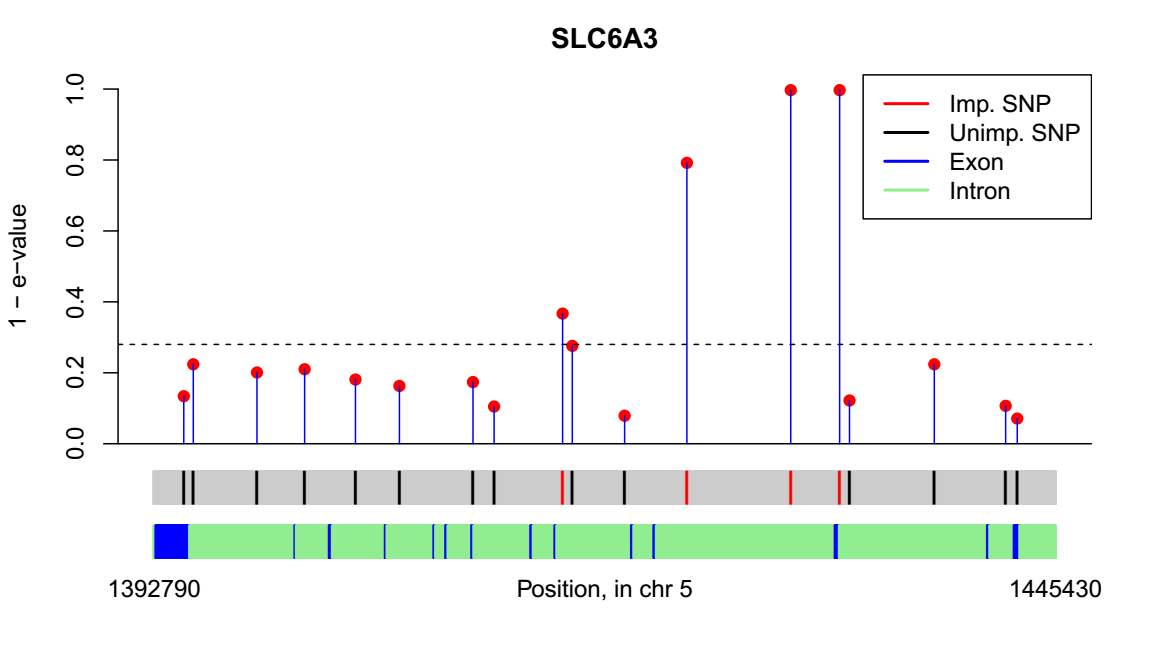}\\
		(g)\\
		\includegraphics[height=.4\textwidth]{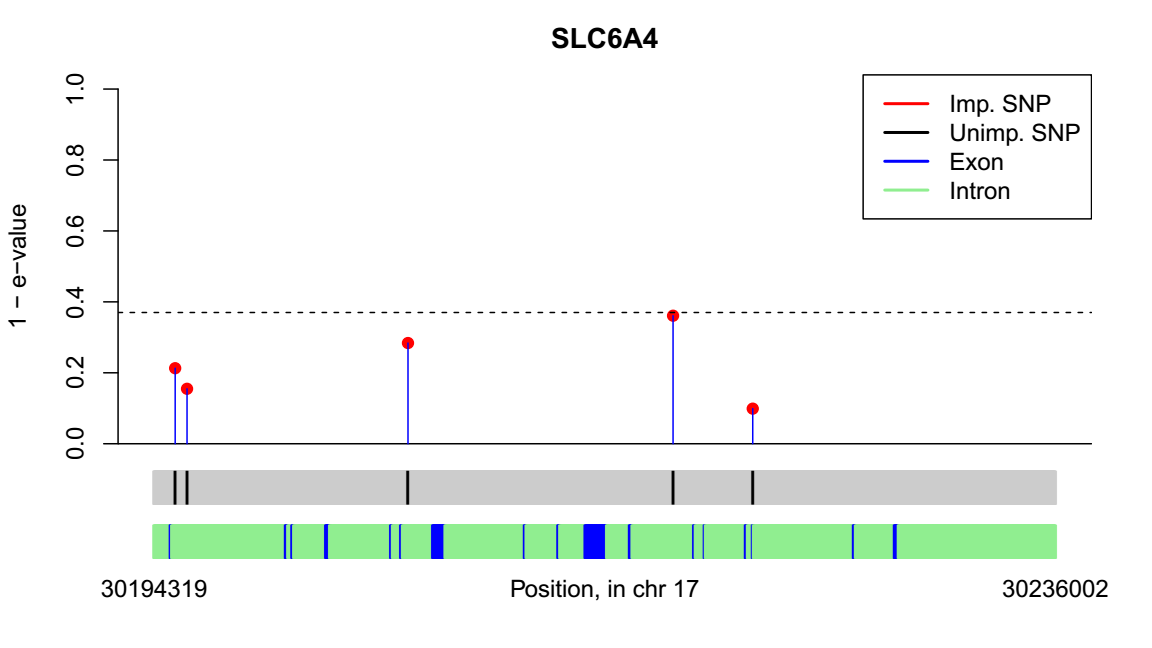} \\
		(h)\\	
		\includegraphics[height=.4\textwidth]{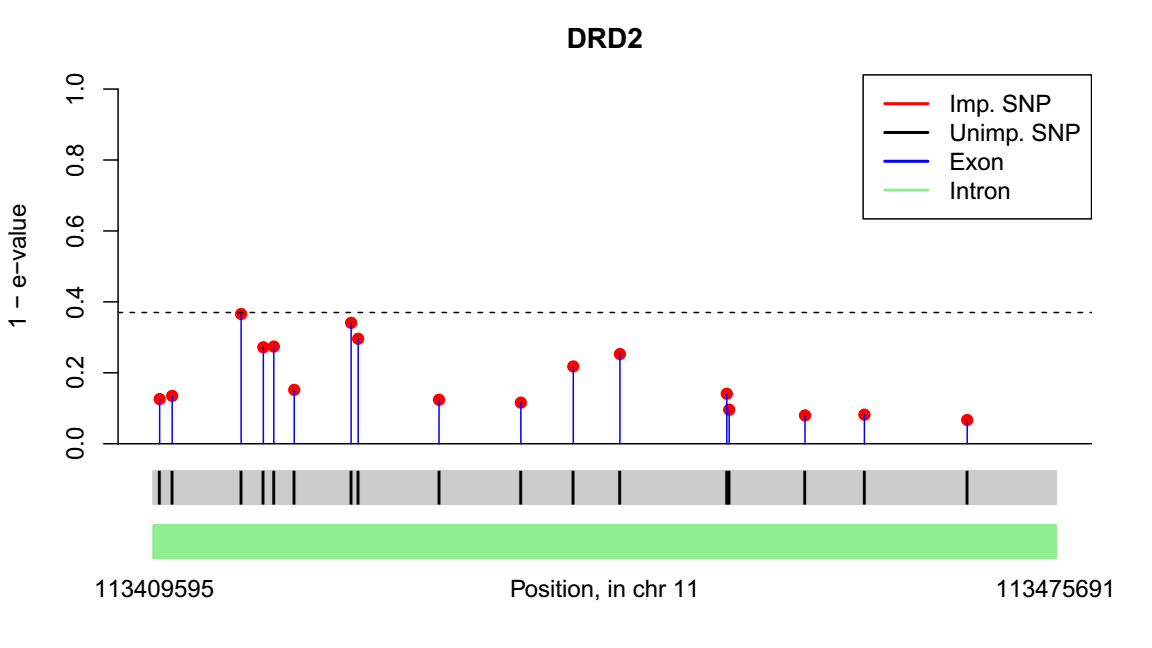}\\
		(i)\\	
\end{tabular}

\caption{Plot of e-values for genes analyzed: (g) SLC6A3, (h) SLC6A4, (i) DRD2}
\label{fig:geneplot3}

\end{center}
\end{figure}

We plot the $90^{\Th}$ quantile e-value estimates in Figures~\ref{fig:geneplot1}, \ref{fig:geneplot2} and \ref{fig:geneplot3}. We obtained gene locations, as well as the locations of coding regions of genes, i.e. exons, inside 6 of these 9 genes from annotation data extracted from the UCSC Genome Browser database \cite{UCSCdata}. Exon locations were not available for OPRM1, CYP2E1 and DRD2. In general, SNPs tend to get selected in groups with neighboring SNPs, which suggests high LD. Also most of the selected SNPs either overlap or in close proximity to the exons, which underline their functional relevance.
\section{Discussion and conclusion}
\label{sec:endSection}


To expand the above approach to a genome-wide scale, we need to incorporate strategies for dealing with the hierarchical structure of pathways and genes: there are only a few genes associated with a quantitative phenotype, which can be further attributed to a small proportion of SNPs inside each gene. To apply the e-values method here, it is plausible to start with an initial screening step to eliminate evidently non-relevant genes. Methods like the grouped Sure Independent Screening \cite{LiZhongZhu12} and min-P test \cite{WestfallYoungBook93} can be useful here. Following this, in a multi-gene predictor set, there are several possible strategies to select important genes \textit{and} important SNPs in them. Firstly, one can use a two-stage e-value based procedure. The first stage is same as the method described in this paper, i.e. selecting important SNPs from each gene using multi-SNP models trained on SNPs in that gene. In the second stage, a model will be trained using the aggregated set of SNPs obtained in the first step, and a group selection procedure will be run on this model using e-values. This means dropping \textit{groups} of predictors (instead of single predictors) from the full model, checking the reduced model e-values, and selecting a SNP group only if dropping it causes the e-value to go below a certain cutoff. Secondly, one can start by selecting important genes using an aggregation method of SNP-trait associations (e.g. \citet{LamparterEtal16}) and then run the e-value based SNP selection on the set of SNPs within these genes. Thirdly, one can also take the aggregated set of SNPs obtained from running the e-values procedure on gene-level models, then use a fast screening method (e.g. RFGLS) to select a subset of those SNPs.

We plan to study merits and demerits of these strategies and the computational issues associated with them in detail through synthetic studies as well as in the GWAS data from MCTFR. Finally, the current evaluation map based formulation requires the existence of an asymptotic distribution for the full model estimate. We plan to explore alternative formulation of evaluation maps under weaker conditions to bypass this, thus being able to tackle high-dimensional ($n < p$) situations.

It is important to remember that in a GWAS setting looking for causal factors of polygenic, quantitative traits is a complex problem. Small effect sizes and high amounts of LD---combined with the influence of environmental covariates---can make finding a set of SNPs behind that trait a noisy process. Typically the random effect error-term is earmarked to account for and quantify such heterogeneities at family-level, but how accurate this quantification is depends on the specific problem context. For this reason, in single-SNP models adjusting the p-values for FDR is important before selecting the final set of SNPs. While our proposed method is based on multi-SNP models, it may still need corrections to calibrate the potential of false discoveries. Finally, robustifying our proposed against data-level issues such as non-nuclear families, lack of individual-level data for some individuals warrant additional research. To this end, there is potential of adapting existing methods, such as \citet{oath}, to the paradigm of e-values.

\section*{Acknowledgements}
This work is a part of the PhD thesis of the first author (SM). He was supported by the University of Minnesota  Interdisciplinary Doctoral Fellowship program during this research. SB was supported by the NIH under grant R01-DA033958. MM was supported by NIH grants R01-AA09367, R01-DA05147 and R01-DA013240 for data collection and genotyping. SC was partially  supported by the National Science Foundation (NSF) under grants 1737918, 1939916 and 1939956.

\section*{Data Availability}
The genotype and phenotype data for the MCTFR sample used in this study are available through the Database of Genotypes and Phenotypes (dbGaP, phs000620.v1.p1). but restrictions apply to the availability of these data, were used under license for the current study, and so are not publicly available. Data are however available from corresponding author on responsible request.

\bibliographystyle{apalike}
\bibliography{Twin-studies-e-values}

\appendix
\section*{Appendix}

\section{Theory of e-values}
\label{sec:methodsSection}
We present the details of our methodology in this Appendix. Sections \ref{subsec:DefineModel} and \ref{subsec:meanEvalue} summarize the existing method of e-values that performs best subset variable selection in a wide range of statistical models \cite{evalues}. We build on this framework and present new results for better detection of weak SNP signals. Section \ref{subsec:BootEvalue} elaborates on the bootstrap implementation of this methodology using the model in \eqref{eqn:LMMeqn}, while Section~\ref{sec:proofs} presents proofs of all theoretical results.

\subsection{Models and evaluation maps}
\label{subsec:DefineModel}
In a general modelling situation where one needs to estimate a set of parameters $\bftheta \in \BR^p$ from an array of samples $\cB_n = \{ B_{1}, \ldots, B_{n} \}$ at stage $n$, any hypothesis or statistical model corresponds to a subset of the full parameter space. Here we consider the model spaces $\bfTheta_{m} \subseteq \BR^{p}$ in which some elements of the parameter vector have fixed values, while others are estimated from the data. Formally, a generic parameter vector $\bftheta_{m} \in \bfTheta_{m}$ consists of entries
$$
\theta_{\cM j} = \left\{ \begin{array}{ll}
\text{ Unknown } \theta_{\cM j} & \text{ for } 
			j \in \cS, \\
\text{ Known } c_{j} & \text{ for } j \notin \cS.
\end{array}
\right.
$$
for some $\cS \subseteq \{ 1, \ldots, p \}$. Thus the estimable index set $\cS$ and fixed elements $C = (c_{j} : j \notin \cS)$ fully specify any model $\cM$ in this setup.


We obtain the full model estimates as minimizers of an estimating equation:
\begin{align}
\label{eqn:EstEqn}
\hat \bftheta = \argmin_{ \bftheta} \Psi (\bftheta) = \argmin_{ \bftheta } \sum_{i=1}^n \psi_i (\bftheta, B_i).
\end{align}
The only condition we impose on these generic estimating functionals $\psi_i(\cdot)$ are:

\noindent\textbf{(P1)} The population version of (\ref{eqn:EstEqn}) has a unique minimizer $\bftheta_0$, i.e.
\begin{align*}
\bftheta_0 = \argmin_{ \bftheta } \BE \sum_{i=1}^n \psi_i (\bftheta, B_i).
\end{align*}
\noindent\textbf{(P2)} There exist a sequence of positive numbers $a_n \uparrow \infty$ and a $p$-dimensional probability distribution $\BT_0$ such that $a_n (\hat \bftheta - \bftheta_0) \leadsto \BT_0$.

We designate $\bftheta_0$ as the {\it true parameter vector}, some elements of which are potentially set to 0. We can now classify any candidate model $\cM$ into one of the two classes: the ones that satisfy $\bftheta_0 \in \bfTheta_\cM$, and the ones that do not. We denote these two types of models by \textit{adequate} and \textit{inadequate models}, respectively. Given the data and unknown $\bftheta_0$, we want to determine if a candidate model is adequate or inadequate.

For this we need coefficient estimates $\hat \bftheta_\cM$ for model $\cM$. We do so by just replacing elements of $\hat \bftheta$ not in $\cS$ by corresponding elements of $\bfc$. This means that for the $j^\text{th}$ element, $j = 1, \ldots, p$, we have
$$
\hat \theta_{\cM j} = \left\{ \begin{array}{ll}
\text{ Unknown } \hat \theta_{j} & \text{ for } 
			j \in \cS, \\
\text{ Known } c_{j} & \text{ for } j \notin \cS.
\end{array}
\right.
$$

We denote the probability distribution of a random variable $\bfT$ by $[\bfT]$. With this notation, we aim to compare the above model estimate distributions with the full model distribution, i.e. $[\hat \bftheta_\cM]$ with  $[\hat\bftheta]$. For this we define an \textit{evaluation map} function $E: \BR^p \times \tilde \BR^p \raro [0, \infty)$ that measures the relative position of $\hat \bftheta_\cM$ with respect to $[\hat \bftheta]$. Here $\tilde \BR^p$ is the set of probability measures on $\BR^p$. We assume that $E$ satisfies the following conditions:

\noindent\textbf{(E1)} For any probability distribution $\BG \in \tilde \BR^p$ and $\bfx \in \BR^p$, $E$ is invariant under location and scale transformations:
$$
E( \bfx, \BG) = E( a\bfx + \bfb, [a \bfG + \bfb]); \quad a \in \BR \neq 0, \bfb \in \BR^p.
$$
where the random variable $\bfG$ has distribution $\BG$.

\noindent\textbf{(E2)} The evaluation map $E$ is lipschitz continuous under the first argument:
$$
| E( \bfx, \BG) - E( \bfy, \BG) | < \| \bfx - \bfy \|^\alpha; \quad \bfx, \bfy \in \BR^p, \alpha > 0.
$$

\noindent\textbf{(E3)}
Suppose $\{ \BY_n \}$ is a tight sequence of probability measures in $\tilde \BR^p$ with weak limit $\BY_\infty$. Then $E( \bfx, \BY_n)$ converges uniformly to $E(\bfx, \BY_\infty)$.

\noindent\textbf{(E4)}
Suppose $\bfZ_n$ is a sequence of random variables such that $\| \bfZ_n \| \stackrel{P}{\raro} \infty$. Then $E( \bfZ_n, \BY_n) \stackrel{P}{\raro} 0$.
\vspace{1em}

For any $\bfx \in \BR^p$ and $[ \bfX ] \in \tilde \BR^p$ with a positive definite covariance matrix $\BV \bfX$, following are examples of the evaluations functions covered by the above set of conditions:
\begin{align}\label{eqn:EvaluationExamples}
E_1 (\bfx, [ \bfX]) = \left[ 1 + \left\| \frac{\bfx - \BE \bfX }{ \sqrt { \diag  (\BV \bfX) }}  \right\|^2 \right]^{-1}; \quad
E_2 (\bfx, [ \bfX]) = \exp \left[ - \left\| \frac{\bfx - \BE \bfX }{ \sqrt { \diag  (\BV \bfX) }}  \right\| \right]
\end{align}
Data depths \cite{tukey75, zuo00, zuo03} also constitute a broad class of point-to-distribution proximity functions that satisfy the above regularity conditions for evaluation maps. Indeed, \citet{evalues} used Mahalanobis depth and halfspace depth \cite{tukey75} as evaluation function to perform model selection. However, the conditions (E1) and (E4) are weaker than those imposed on a traditional depth function \cite{zuo00}. Conditions (E2) and (E3) are not required of depth functions in general, but they arise implicitly in several implementations of data depth \cite{MoslerChapter13}. The theoretical results we state here are based on a general evaluation map and not depth functions. To emphasize this point, in this paper we use the non-depth evaluation functions $E_1$ and $E_2$ as in (\ref{eqn:EvaluationExamples}) above.

\subsection{Model selection using e-values}
\label{subsec:meanEvalue}
Depending on the choice of the data sequence $\cB_n$, $E( \hat \bftheta_\cM, [\hat \bftheta])$ can take different values. For any candidate model $\cM$, we denote the distribution of the corresponding evaluation map---across all possible random data sequences $\cB_n$---by $\BE_{\cM n}$. For simplicity we drop the $n$ in its subscript, i.e. $\BE_{\cM n} \equiv \BE_\cM $. These distributions are informative of the behavior of parameter estimates $\hat \theta_\cM$. We use them as a tool to distinguish between inadequate and adequate models. Given a single set of samples, we use resampling schemes that satisfy standard regularity conditions \cite{evalues} to get consistent approximations of $\BE_\cM$.

We now define the \textbf{e-value} to compare the different model estimates and eventually perform selection of important SNPs from a multi-SNP model. Loosely construed, any functional of the evaluation map distribution $\BE_\cM$ that can act as model evidence is an e-value. For example, \citet{evalues} took the mean functional of $\BE_\cM$ (say $\mu (\BE_{\cM })$) as e-value, and proved the following result (see Theorem 5.1 therein):

\begin{Theorem}\label{thm:meanThm}
Consider estimators satisfying conditions (P1) and (P2), and an evaluation map $E$ satisfying the conditions (E1), (E2) and (E4). Also suppose that
$$
\lim_{n \raro \infty} \mu ( \BY_n ) = \mu ( \BY_\infty) < \infty,
$$
for any tight sequence of probability measures $\{ \BY_n \}$ in $\tilde \BR^p$ with weak limit $\BY_\infty$. Then as $n \raro \infty$,

\begin{itemize}
\item For the full model, $\mu (\BE_{* }) \raro \mu_\infty$ for some $0 < \mu_\infty < \infty$;
\item For any adequate model, $| \mu ( \BE_\cM ) - \mu (\BE_{* }) | \raro 0$,
\item For any inadequate model, $ \mu ( \BE_\cM) \raro 0$.
\end{itemize}
\end{Theorem}

Taking data depths as evaluation functions leads to a further result that $\mu (\BE_{* }) < \mu (\BE_\cM)$ for any adequate model $\cM$ and large enough $n$.

A similar result holds for tail quantiles as well. Denote the $q^{\Th}$ population quantile of $\BE_\cM$ by $c_{q} ( \BE_\cM ) $. Then we have equivalent results to Theorem \ref{thm:meanThm} as $n \raro \infty$:

\begin{Theorem}
\label{thm:quantileThm}
Given that the estimator $\hat \bftheta$ satisfies conditions (P1) and (P2), and the evaluation map satisfies conditions (E1)-(E4), we have
\begin{align}
& c_{q} ( \BE_{* } ) \raro c_{q,\infty} < \infty, \label{evalEqn1}\\
& | c_{q} ( \BE_\cM ) - c_{q} ( \BE_{* } ) | \raro 0 \text{ when } \cM \text{ is adequate},\label{evalEqn2}\\
& c_{q} ( \BE_\cM ) \raro 0 \text{ when } \cM \text{ is inadequate}. \label{evalEqn3}
\end{align}
\end{Theorem}
A one-step model selection procedure equivalent to Algorithm 1 follows when we restrict our model selection procedure to the restricted model class
$$ \BM_0  = \{ \cM \equiv (\cS, C): c_j = 0 \quad\forall\quad j \notin \cS \}. $$
In other words, we only consider models that have known parameters set at
0. In this scenario, non-zero indices of $\bftheta_0$ (say $\cS_0$) can be recovered through Algorithm 1. This follows from applying Corollary 5.3 in \citet{evalues} in our setup, with mean e-values replaced by quantile e-values.

\subsection{Bootstrap procedure}
\label{subsec:BootEvalue}
We use GBS~\cite{ChatterjeeBose05} to obtain approximations of the sampling distributions $\BE_{-j}$ and $\BE_*$. It calculates bootstrap equivalents of the parameter estimate $\hat \bftheta$ by minimizing a version of the estimating equation in (\ref{eqn:EstEqn}) with random weights:
\begin{align}
\hat \bftheta_r = \argmin_{ \bftheta} \sum_{i=1}^n \BW_{i} \psi_i  ( \bftheta, B_i).
\end{align}
The resampling weights $(\BW_{1}, \ldots, \BW_{n})$ are non-negative exchangeable random variables chosen independent of the data, and satisfy the following conditions:
\begin{eqnarray}
\BE \BW_1  = 1; \quad
\BV \BW_1  = \tau_{n}^{2} \uparrow \infty; \quad
\tau_n^{2} 
= o ( a_n^{2})\label{eq:ResamplingweightBW}\\
\BE W_{1} W_{2} = O (n^{-1}); \quad
\BE W_{1}^{2} W_{2}^{2} \raro 1; \quad
\BE W_{1}^{4} < \infty \label{eq:ResamplingweightW}
\end{eqnarray}
with $W_i := (\BW_i - 1)/ \tau_n; i = 1, \ldots, n$ being the centered and scaled resampling weights. Under standard regularity conditions on the estimating functional $\Psi(\cdot)$ \cite{ChatterjeeBose05, evalues} and conditional on the data, $(a_n/\tau_n) (\hat \bftheta_w - \hat \bftheta)$ converges to the same asymptotic distribution as $a_n ( \hat \bftheta - \bftheta_0 )$, i.e. $\BT_0$.

We use empirical quantiles of the full model bootstrap samples as the quantile e-value estimates. Specifically, we go through the following steps:
\begin{itemize}
\item Fix $q,t \in (0,1)$;

\item Generate two independent set of bootstrap weights, of size $R$ and $R_1$, and obtain the corresponding approximations to the full model sampling distribution, say $[ \hat\bftheta_{r} ]$ and $[ \hat\bftheta_{r_1} ]$;

\item For $j = 1,2,\ldots p$ and estimate the e-value of the $j^{\Th}$ predictor as the empirical $q^{\Th}$ quantile of $\hat \BE_{-j} := [ E( \hat \bftheta_{r,-j}, [ \hat \bftheta_{r_1}]) ]$, with $\hat \bftheta_{r,-j}$ obtained from $\hat \bftheta_{r}$ by replacing the $j^{\Th}$ coordinate with 0;

\item Estimate the set of non-zero covariates as 
$ \hat \cS_{0} = \lbrace j: c_q (\hat\BE_{-j}) <  c_{q t} (\hat\BE_{*}) \} $.
\end{itemize}
Conditions (\ref{eq:ResamplingweightBW}) and (\ref{eq:ResamplingweightW}) on the resampling weights ensure bootstrap-consistent approximation of the evaluation map quantiles:

\begin{Theorem}
\label{thm:bootThm}
Given the estimator $\hat \bftheta$ and evaluation map $E$ in Theorem~\ref{thm:quantileThm}, and a generalized bootstrap scheme satisfying (\ref{eq:ResamplingweightBW}) and (\ref{eq:ResamplingweightW}), we get
\begin{align}
& | c_{q} ( \hat \BE_\cM ) - c_{q} ( \hat \BE_{* } ) | \stackrel{P_n}{\raro} o_P(1) \text{ when } \cM \text{ is adequate},\label{bootevalEqn2}\\
& c_{q} ( \hat \BE_\cM ) \stackrel{P_n}{\raro} o_P(1) \text{ when } \cM \text{ is inadequate}. \label{bootevalEqn3}
\end{align}
where $P_n$ is probability conditional on the data.
\end{Theorem}

Generalized bootstrap covers a large array of resampling procedures, for example the $m$-out-of-$n$ bootstrap and a scale-enhanced version of the bayesian bootstrap. 
 Furthermore, given that $\psi_i(.)$ are twice differentiable in a neighborhood of $\bftheta_0$ and some other conditions in \citet{ChatterjeeBose05}, there is an approximate representation of $\hat \bftheta_w$:
\begin{align}\label{eqn:BootEqn}
\hat\bftheta_r = \hat\bftheta - \frac{ \tau_n}{a_n} \left[ \sum_{i=1}^n W_i \psi_i'' (\hat \bftheta, B_i) \right]^{-1}
\sum_{i=1}^n W_i \psi_i' (\hat \bftheta, B_i) + \bfR_{wn}.
\end{align}
with $\BE_w \| \bfR_{w n} \|^2 = o_P(1)$.

Given the full model estimate $\hat \bftheta$, and the score vectors $ \psi_i' (\hat \bftheta, B_i)$ and hessian matrices $ \psi_i'' (\hat \bftheta, B_i)$, (\ref{eqn:BootEqn}) allows us to obtain multiple copies of $\hat \bftheta_r$ through Monte-Carlo simulation of several arrays of bootstrap weights. This bypasses the need to fit the full model for each bootstrap sample, resulting in extremely fast computation of e-values.

We adapt the approximation of (\ref{eqn:BootEqn}) to the LMM in (\ref{eqn:LMMeqn}). We first obtain the maximum likelihood estimates $\hat \bfbeta_g, \hat \sigma_a^2, \hat \sigma_c^2, \hat \sigma_e^2$ through fitting the LMM. Then we replace the variance components in (\ref{eqn:partsOfV}) with corresponding estimates to get $\hat \bfV_i$ for $i^{\Th}$ pedigree, and aggregate them to get the covariance matrix estimate for all samples:
$$
\hat \bfV = \diag ( \hat \bfV_1, \ldots, \hat \bfV_m ).
$$
We take $m$ random draws from Gamma$(1,1) - 1$, say $\{ w_{r1}, \ldots, w_{rm} \}$, as resampling weights in (\ref{eqn:BootEqn}), using the same weight for all members of a pedigree. Consequently, the bootstrapped coefficient estimate $\hat \bfbeta_{r g}$ has the following representation:
\begin{align}
\hat \bfbeta_{r g} & \simeq \hat \bfbeta_g + \frac{ \tau_n}{\sqrt n} ( \bfG^T \hat \bfV^{-1} \bfG )^{-1}
\bfW_r \bfG^T \hat \bfV^{-1} ( \bfy - \bfG \hat \bfbeta_g).
\end{align}
with $\bfG = ( \bfG_1^T, \ldots, \bfG_m^T )^T$ and $\bfW_r = \diag (w_{r1} \bfI_4, \ldots, w_{r m} \bfI_4 )$. Finally we repeat the procedure for two independent sets of resampling weights, say of sizes $R$ and $R_1$, to obtain two collections of bootstrapped estimates $\{ \hat \bfbeta_{1g}, \ldots, \hat \bfbeta_{Rg} \}$.

\subsection{Proof of theoretical results}
\label{sec:proofs}

\begin{proof}[Proof of Theorem~\ref{thm:quantileThm}]
Define $c_{q,\infty} = q^{\text{th}}$ quantile of $\BT_0$. Now following assumption (E1),
\begin{align*}
c_q (\BE_*) 
&= \inf_\bftheta \{ E(\bftheta, [\hat \bftheta]): \BF_* \geq q \} \\
&= \inf_\bftheta \{ E(a_n(\bftheta - \bftheta_0), [a_n(\hat \bftheta - \bftheta_0)]):  a_n (\BF_* - \bftheta_0) \geq q\}.
\end{align*}
where $\BF_*$ is the probability distribution function of $E(\hat \bftheta, [\hat \bftheta])$. Part 1 is proved following assumptions (P2) and (E3).

Now if $\cM$ is adequate, following assumption (E1),
\begin{align}\label{eqn:Thm1ProofEqn1}
E (\hat \bftheta_\cM, [\hat \bftheta]) = E( \hat \bftheta_\cM - \bftheta_0, [\hat \bftheta - \bftheta_0]).
\end{align}
Decompose the first argument as
\begin{align}\label{eqn:Thm1ProofEqn2}
\hat \bftheta_\cM - \bftheta &=
(\hat \bftheta_\cM - \hat \bftheta) + (\hat \bftheta - \bftheta_0).
\end{align}
By definition, $\hat \theta_{\cM j} - \hat \theta_j = 0$ if $j \in \cS$, else equals $\theta_{0j} - \hat \theta_j$. Thus for the first summand in (\ref{eqn:Thm1ProofEqn2}) we have
$$
\hat \bftheta_\cM - \hat \bftheta =  O_P (1/a_n).
$$
Going back to (\ref{eqn:Thm1ProofEqn1}), this implies
$$
| E( \hat \bftheta_\cM - \bftheta_0, [\hat \bftheta - \bftheta_0]) -
E( \hat \bftheta - \bftheta_0, [\hat \bftheta - \bftheta_0]) | < O_P (a_n^{-\alpha}),
$$
using lipschitz continuity in assumption (E2), i.e
$$
| E( \hat \bftheta_m, [\hat \bftheta]) - E( \hat \bftheta, [\hat \bftheta]) | < O_P (a_n^{-\alpha}),
$$
again using (E1). Part 2 now follows.

For part 3, we apply (E1) to get
\begin{align}\label{eqn:Thm1ProofEqn3}
E (\hat \bftheta_m, [\hat \bftheta])
= E( a_n(\hat \bftheta_\cM - \bftheta_0), [a_n(\hat \bftheta - \bftheta_0)]).
\end{align}
We then decompose the first argument as
\begin{align}
\label{eqn:Thm1ProofEqn4}
a_n(\hat \bftheta_\cM - \bftheta_0)
&= a_n(\hat \bftheta_\cM - \bftheta_\cM) + a_n(\bftheta_\cM - \bftheta_0)
\end{align}
Since $\cM$ is inadequate, $\theta_{\cM j} \neq \theta_{0j}$ when $j \notin \cS$. So $\| a_n(\bftheta_\cM - \bftheta_0) \| \uparrow \infty$ as $a_n \uparrow \infty$. Applying (E4) now proves part 3.
\end{proof}

\begin{proof}[Proof of Theorem~\ref{thm:bootThm}]
The proof is fairly similar to that of Theorem 5.1 in \citet{evalues}, so we give a sketch. For the full model, the bootstrap is consistent, i.e. $a_n (\hat \bftheta_* - \bftheta_0)$ and $(a_n/ \tau_n) (\hat \bftheta_{r*} - \hat \bftheta_*)$ converge to same weak limit in probability, following theorems 2.2 and 2.3 in \citet{evalues}. Specifically, conditions (A1)-(A6) in \citet{evalues} ensure condition (P2) in our paper through theorem 2.2 therein, following which  theorem 2.3 ensures that when (A1)-(A6) are satisfied, bootstrap consistency holds. The definition of $\hat \bftheta_\cM$ now means that
$a_n (\hat \bftheta_\cM - \bftheta_\cM)$ and $(a_n/ \tau_n) (\hat \bftheta_{r\cM} - \hat \bftheta_\cM)$ converge to the same weak limit in probability for any model $\cM$. A similar approach as the proof of parts 2 and 3 of theorem 5.1 now follows, with an additional term corresponding to bootstrap estimates in (\ref{eqn:Thm1ProofEqn2}) and (\ref{eqn:Thm1ProofEqn4}).
\end{proof}

\section{Discussion on gene-specific findings in the MCTFR data}
\label{sec:DataSection}

{\it GABRA2:} As seen in the plots, the first two SNPs detected are close to two separate exons. The 4th and 5th detected SNPs, rs1808851 and rs279856, are at perfect LD with rs279858 in the larger 7188-individual dataset \cite{IronsThesis12}. This SNP had not been genotyped in our sample, but is the marker in GABRA2 that is most frequently associated in the literature with alcohol abuse \cite{CuiEtal12}. Interestingly, a single SNP RFGLS analysis of the same twin studies data that used Bonferroni correction on marginal $p$-values to detect SNPs had missed these SNPs \cite{IronsThesis12}. This highlights the advantage of our approach.

{\it ADH genes:} Multiple studies have associated rs1229984 in the ADH1B gene (position 99318162 of chromosome 4) with alcohol dependence (\url{https://www.snpedia.com/index.php/Rs1229984}), which as seen in the plot of ADH2 is close to an exon region. Our data does not contain this marker, but detects one SNP 20 kb upstream of this, rs17027523. Another SNP, rs3775540 at position 99304544 has an $e$-value of 0.226, so narrowly misses detection. This is close to rs1229984, and also rs1042026 at position 99307309, which \citet{MacgregorEtal08} found to be strongly associated with alcohol consumption.

The SNP rs17027523 is interesting: it resides in the uncharacterized long non-coding RNA gene LOC100507053. One previous study \cite{GelernterEtal14, XuEtal15} found significant associations for 5 SNPs in this gene with alcohol consumption for African American population through single-SNP analysis on non-familial GWAS data. Notably, their analysis found a much stronger evidence of the association in African-American part of the sample than the European American part, while our findings are entirely from a Caucasian sample.

{\it OPRM1:} Many of the SNPs analyzed in this gene have very low $e$-values, and tend to cluster together. The minor allele of the SNP rs1799971 (chr 6, position 154039662) has been associated with stronger alcohol cravings (\url{https://www.snpedia.com/index.php/Rs1799971}), and we detect rs12662873 at position 154040810.

{\it CYP2E1:} Five of the 9 SNPs studied are detected through our analysis. Four of them are within 10 kb of one another (base pairs 133534822 to 133543210 in chr 10). In the analysis of \citet{LindEtal12} rs4646976 at 133534223 position was most associated with a measure of breath alcohol concentration: this is within our detected region. This study had also detected rs4838767 in the promoter region of CYP2E1 (position 133520114) associated with multiple alcohol consumption measures. We detect rs9419702 at position 133531153.

{\it ALDH2:} All 6 SNPs we study are close to exons, and 5 get picked up by the $e$-value procedure. While all five are at a lesser base pair position than the well-known SNP rs671 ({\url{https://www.snpedia.com/index.php/Rs671}, position 111803962), one of the SNPs we analyze (rs16941437) is within 10 kb upstream of this SNP.

{\it COMT:} The SNP rs4680 has long been associated with schizophrenia and substance abuse, including alcoholism. A case-control study \cite{VoiseyEtal11} associated rs4680 and rs165774 with alcohol dependence through a SNP-wise chi-squared test, and had these two SNPs in high LD in their study population. Compared to this, in our simultaneous model of all COMT polymorphisms, the more well-known rs4680 has a below threshold $e$-value. 

{\it SLC6A3:} Our analysis does not detect rs27072, which has been associated with alcohol withdrawal symptoms (\url{https://www.snpedia.com/index.php/Rs27072}). 
%
%

Finally, most $e$-values for the last 3 genes, i.e. SLC6A3, SLC6A4 and DRD2, are large: indicating weak SNP signals. We found this observation interesting, because variants of these genes have known interaction effects behind alcohol withdrawal-induced seizure \cite{KarpyakEtal10} and bipolar disorder \cite{WangEtal14}, as well as additive effect on the susceptibility to smoking addiction \cite{ErblichEtal05}.

 

\section{Outputs for MCTFR data analysis}
\label{sec:SnpSection}
Each table gives the $90^{\text{th}}$ percentile $e$-values, which are plotted in figures 2, 3, and 4 in main paper, of SNPs analyzed in the gene. Column 'Association' is obtained from the sign of the SNP coefficient in the full model.

\begin{table}[ht]
\centering
\begin{tabular}{lccc}
  \hline
SNP name & Location & $e$-value & Association \\ 
  \hline
rs16859227 & 46250605 & 0.89 & + \\ 
  rs572227 & 46251393 & 0.13 & - \\ 
  rs534459 & 46256805 & 0.24 & + \\ 
  rs2119183 & 46272806 & 0.92 & - \\ 
  rs502038 & 46280318 & 0.58 & + \\ 
  rs1808851 & 46311447 & 0.00 & + \\ 
  rs279856 & 46317923 & 0.00 & - \\ 
  rs3775282 & 46321863 & 0.86 & - \\ 
  rs279841 & 46340763 & 0.75 & + \\ 
  rs10805145 & 46358331 & 0.73 & - \\ 
  rs13152740 & 46381221 & 0.86 & - \\ 
   \hline
\end{tabular}
\caption{SNPs for GABRA2, chr4, position 46243548 - 46390039; $e$-value cutoff 0.72}
\end{table}

\begin{table}[ht!]
\centering
\begin{tabular}{lccc}
  \hline
SNP name & Location & $e$-value & Association \\ 
  \hline
rs17027299 & 99078105 & 0.84 & - \\ 
  rs9307222 & 99101051 & 0.76 & - \\ 
  rs10006414 & 99101401 & 0.49 & + \\ 
  rs9994641 & 99101605 & 0.48 & + \\ 
  rs13134014 & 99104879 & 0.75 & - \\ 
  rs6820691 & 99105055 & 0.76 &+\\ 
  rs6820913 & 99125659 & 0.81 &+\\ 
  rs6532729 & 99146436 & 0.67 &-\\ 
  rs13150538 & 99152631 & 0.49 &-\\ 
  rs17027380 & 99157450 & 0.63 &-\\ 
  rs17494998 & 99160699 & 0.41 &+\\ 
  rs549467 & 99172232 & 0.81 &+\\ 
  rs2034677 & 99187874 & 0.62 &+\\ 
  rs12508445 & 99190653 & 0.01 &-\\ 
  rs10003496 & 99197839 & 0.81 &+\\ 
  rs10005811 & 99208603 & 0.02 &+\\ 
  rs603215 & 99214851 & 0.78 &-\\ 
  rs433146 & 99229839 & 0.87 &-\\ 
  rs17027456 & 99235747 & 0.31 &-\\ 
  rs17561798 & 99235941 & 0.85 &+\\ 
  rs10516428 & 99237439 & 0.45 &-\\ 
  rs6532731 & 99251006 & 0.80 &+\\ 
  rs7694221 & 99260423 & 0.90 &+\\ 
  rs10028330 & 99268949 & 0.70 &-\\ 
  rs10022047 & 99296818 & 0.49 &+\\ 
  rs17027523 & 99298979 & 0.05 &+\\ 
  rs17027530 & 99303633 & 0.69 &+\\ 
  rs3775540 & 99304544 & 0.23 &-\\ 
  rs3756088 & 99309404 & 0.89 &-\\ 
  rs13103626 & 99317251 & 0.75 &+\\ 
  rs10516430 & 99337881 & 0.62 &+\\ 
  rs9884594 & 99359318 & 0.68 &-\\ 
  rs12503056 & 99369061 & 0.63 &+\\ 
  rs2004316 & 99381148 & 0.43 &-\\ 
  rs4303985 & 99399748 & 0.87 &-\\ 
  rs4414961 & 99403784 & 0.86 &-\\ 
  rs12509267 & 99407299 & 0.80 &+\\ 
  rs6838913 & 99408106 & 0.84 &-\\ 
  rs4374629 & 99411783 & 0.85 &+\\ 
  rs4527483 & 99421741 & 0.89 &+\\ 
  rs10009693 & 99423280 & 0.90 &-\\ 
  rs10023791 & 99425353 & 0.88 &+\\ 
  rs955931 & 99428163 & 0.88 &-\\ 
  rs17027628 & 99428608 & 0.85 &-\\ 
   \hline
\end{tabular}
\caption{SNPs for ADH genes, chr4, position 99070977 - 99435737; $e$-value cutoff 0.225}
\end{table}

\begin{table}[ht]
\centering
\begin{tabular}{lccc}
  \hline
SNP name & Location & $e$-value & Association \\ 
  \hline
rs2000371 & 154011024 & 0.39 &-\\ 
  rs9371718 & 154011615 & 0.08 &-\\ 
  rs12211203 & 154016936 & 0.63 &-\\ 
  rs1937600 & 154017197 & 0.02 &-\\ 
  rs9397637 & 154022718 & 0.00 &+\\ 
  rs1937590 & 154036895 & 0.63 &+\\ 
  rs12662873 & 154040810 & 0.18 &+\\ 
  rs12661209 & 154044112 & 0.84 &-\\ 
  rs1316368 & 154055754 & 0.00 &-\\ 
  rs1937587 & 154060023 & 0.27 &-\\ 
  rs6921403 & 154063906 & 0.00 &-\\ 
  rs1937580 & 154076643 & 0.00 &+\\ 
  rs1937645 & 154082228 & 0.00 &+\\ 
  rs1892361 & 154099619 & 0.00 &-\\ 
  rs1937633 & 154104857 & 0.04 &-\\ 
  rs1937631 & 154105011 & 0.00 &-\\ 
  rs12527197 & 154107836 & 0.02 &+\\ 
  rs1892360 & 154111701 & 0.74 &-\\ 
  rs1892359 & 154112042 & 0.65 &-\\ 
  rs1892356 & 154112263 & 0.56 &+\\ 
  rs1937622 & 154113139 & 0.54 &-\\ 
  rs10485258 & 154113409 & 0.72 &-\\ 
  rs1937619 & 154114583 & 0.58 &-\\ 
  rs1748289 & 154121980 & 0.77 &-\\ 
  rs1781619 & 154135968 & 0.64 &-\\ 
  rs652051 & 154139344 & 0.74 &+\\ 
  rs10485262 & 154140199 & 0.69 &-\\ 
  rs9371312 & 154145492 & 0.81 &+\\ 
  rs1332849 & 154151117 & 0.48 &-\\ 
  rs9371749 & 154153369 & 0.28 &+\\ 
  rs9285539 & 154154532 & 0.08 &+\\ 
  rs9322439 & 154156250 & 0.07 &+\\ 
  rs11752884 & 154159710 & 0.25 &-\\ 
  rs4869813 & 154173845 & 0.13 &+\\ 
  rs4870241 & 154174963 & 0.00 &-\\ 
  rs9384156 & 154186720 & 0.13 &+\\ 
  rs2065139 & 154192175 & 0.89 &-\\ 
  rs689219 & 154198820 & 0.00 &-\\ 
  rs9371761 & 154202578 & 0.20 &-\\ 
  rs12199858 & 154204327 & 0.00 &+\\ 
  rs9371762 & 154213973 & 0.00 &-\\ 
  rs612450 & 154214357 & 0.00 &-\\ 
  rs9384159 & 154219177 & 0.00 &+\\ 
  rs6938958 & 154220427 & 0.00 &-\\ 
  rs581564 & 154221214 & 0.00 &+\\ 
  rs12202611 & 154237443 & 0.76 &-\\ 
  rs4870255 & 154237937 & 0.88 &-\\ 
   \hline
\end{tabular}
\caption{SNPs for OPRM1, chr6, position 154010496 - 154246867; $e$-value cutoff 0.225}
\end{table}

\begin{table}[ht]
\centering
\begin{tabular}{lccc}
  \hline
SNP name & Location & $e$-value & Association \\ 
  \hline
rs10872828 & 133525348 & 0.72 &-\\ 
  rs9419702 & 133531153 & 0.09 &-\\ 
  rs7083395 & 133532269 & 0.77 &+\\ 
  rs9419624 & 133534822 & 0.06 &+\\ 
  rs7906770 & 133536902 & 0.28 &-\\ 
  rs9419569 & 133541881 & 0.06 &+\\ 
  rs9419629 & 133543210 & 0.06 &+\\ 
  rs7093241 & 133556596 & 0.72 &-\\ 
  rs9419649 & 133561098 & 0.91 &-\\ 
   \hline
\end{tabular}
\caption{SNPs for CYP2E1, chr10, position 133520406 - 133561220; $e$-value cutoff 0.72}
\end{table}

\begin{table}[ht]
\centering
\begin{tabular}{lccc}
  \hline
SNP name & Location & $e$-value & Association \\ 
  \hline
rs7398343 & 111774068 & 0.34 &-\\ 
  rs7297186 & 111778178 & 0.36 &+\\ 
  rs3803167 & 111785586 & 0.00 &+\\ 
  rs10219736 & 111788402 & 0.00 &-\\ 
  rs16941437 & 111793039 & 0.00 &-\\ 
  rs3742004 & 111798553 & 0.75 &+\\ 
   \hline
\end{tabular}
\caption{SNPs for ALDH2, chr12, position 111766887 - 111817529; $e$-value cutoff 0.72}
\end{table}

\begin{table}[h!]
\centering
\begin{tabular}{lccc}
  \hline
SNP name & Location & $e$-value & Association \\ 
  \hline
rs4646312 & 19948337 & 0.41 & -\\ 
  rs165656 & 19948863 & 0.22 & -\\ 
  rs165722 & 19949013 & 0.24 &    +\\ 
  rs2239393 & 19950428 & 0.50 &+\\ 
  rs4680 & 19951271 & 0.60 &+\\ 
  rs4646316 & 19952132 & 0.81 &-\\ 
  rs165774 & 19952561 & 0.72 &-\\ 
  rs174699 & 19954458 & 0.07 &+\\ 
  rs165599 & 19956781 & 0.58 &-\\ 
  rs165728 & 19957023 & 0.02 &-\\ 
  rs165815 & 19959473 & 0.00 &+\\ 
  rs5993891 & 19959746 & 0.04 &-\\ 
  rs887199 & 19961955 & 0.04 &-\\ 
  rs2239395 & 19962203 & 0.07 &+\\ 
  rs2518824 & 19962963 & 0.59 &+\\ 
   \hline
\end{tabular}
\caption{SNPs for COMT, chr22, position 19941607 - 19969975; $e$-value cutoff 0.72}
\end{table}

\begin{table}[h!]
\centering
\begin{tabular}{lrrr}
  \hline
SNP name & Location & $e$-value & Association \\ 
  \hline
rs27072 & 1394522 & 0.87 &+\\ 
  rs40184 & 1395077 & 0.78 &-\\ 
  rs11564771 & 1398797 & 0.80 &-\\ 
  rs11133767 & 1401580 & 0.79 &+\\ 
  rs6869645 & 1404548 & 0.82 &+\\ 
  rs3776512 & 1407116 & 0.84 &+\\ 
  rs6347 & 1411412 & 0.83 &-\\ 
  rs27048 & 1412645 & 0.90 &-\\ 
  rs2042449 & 1416646 & 0.63 &+\\ 
  rs13161905 & 1417212 & 0.72 &-\\ 
  rs2735917 & 1420268 & 0.92 &+\\ 
  rs464049 & 1423905 & 0.21 &-\\ 
  rs460700 & 1429969 & 0.00 &-\\ 
  rs460000 & 1432825 & 0.00 &+\\ 
  rs4975646 & 1433401 & 0.88 &-\\ 
  rs403636 & 1438354 & 0.78 &-\\ 
  rs2617605 & 1442521 & 0.89 &+\\ 
  rs6350 & 1443199 & 0.93 &+\\ 
   \hline
\end{tabular}
\caption{SNPs for SLC6A3, chr5, position 1392790 - 1445430; $e$-value cutoff 0.72}
\end{table}

\begin{table}[ht]
\centering
\begin{tabular}{lccc}
  \hline
SNP name & Location & $e$-value & Association \\ 
  \hline
rs16967029 & 30195292 & 0.79 &+\\ 
  rs2051810 & 30195841 & 0.84 &-\\ 
  rs11658318 & 30206059 & 0.72 &-\\ 
  rs8079471 & 30218317 & 0.64 &+\\ 
  rs3760454 & 30222002 & 0.90 &+\\ 
   \hline
\end{tabular}
\caption{SNPs for SLC6A4, chr17, position 30194319 - 30236002; $e$-value cutoff 0.63}
\end{table}

\begin{table}[ht]
\centering
\begin{tabular}{lccc}
  \hline
SNP name & Location & $e$-value & Association \\ 
  \hline
rs2514229 & 113410000 & 0.87 &-\\ 
  rs11214654 & 113410917 & 0.86 &+\\ 
  rs7937641 & 113415976 & 0.63 &-\\ 
  rs12222458 & 113417603 & 0.73 &-\\ 
  rs10736470 & 113418371 & 0.73 &-\\ 
  rs12576506 & 113419869 & 0.85 &+\\ 
  rs10750025 & 113424042 & 0.66 &+\\ 
  rs7952106 & 113424558 & 0.70 &-\\ 
  rs4373974 & 113430486 & 0.88 &-\\ 
  rs4130345 & 113436487 & 0.88 &-\\ 
  rs7123697 & 113440331 & 0.78 &+\\ 
  rs6589386 & 113443753 & 0.75 &+\\ 
  rs4132966 & 113451589 & 0.86 &+\\ 
  rs7940164 & 113451765 & 0.90 &-\\ 
  rs4245155 & 113457324 & 0.92 &-\\ 
  rs11607834 & 113461680 & 0.92 &-\\ 
  rs12280220 & 113469219 & 0.93 &-\\ 
   \hline
\end{tabular}
\caption{SNPs for DRD2, chr11, position 113409595 - 113475691; $e$-value cutoff 0.63}
\end{table}

\end{document}